\documentstyle[prl,aps]{revtex}
\begin{document}
\input psfig

\title{Lattice knot theory and quantum gravity 
in the loop representation}

\author{Hugo Fort, Rodolfo Gambini}
\address{Instituto de F\'{\i}sica, Facultad de Ciencias, \\
Tristan Narvaja 1674, Montevideo, Uruguay}

\author{Jorge Pullin}
\address{
Center for Gravitational Physics and Geometry, Physics Department,\\
The Pennsylvania State University, 104 Davey Lab, University Park, PA 16802}

\maketitle
\begin{abstract}
We present an implementation of the loop representation of quantum gravity
on a square lattice. Instead of starting from a classical lattice theory,
quantizing and introducing loops, we proceed backwards, setting up 
constraints in the lattice loop representation and showing that they 
have appropriate (singular) continuum limits and algebras. The diffeomorphism
constraint reproduces the classical algebra in the continuum and has as
solutions lattice analogues of usual knot invariants. We discuss some of 
the invariants stemming from Chern--Simons theory in the lattice context,
including the issue of framing. We also present a regularization of the
Hamiltonian constraint. We show that  two knot invariants from Chern--Simons
theory are annihilated by the Hamiltonian constraint through the use of 
their skein relations, including intersections. We also discuss the issue
of intersections with kinks. This paper is the first step towards setting
up the loop representation in a rigorous, computable setting. 
\end{abstract}
\vspace{-10cm} 
\begin{flushright}
\baselineskip=15pt
gr-qc/9608033\\
CGPG-96/8-1  \\
ESI-368\\
\end{flushright}
\vspace{8cm}

\section{Introduction}

The use of lattice techniques has proved very fruitful in gauge
theories. This is mainly due to two reasons: lattices provide a
regularization procedure compatible with gauge invariance and they allow
to put the theory on a computer and to calculate observable quantities. 
Based on these successes one may be tempted to apply these techniques to
quantum gravity. There the situation is more problematic. Lattices
introduce preferred directions in spacetime and therefore break the
gauge symmetry of the theory, diffeomorphism invariance. If one studies
a canonical formulation of quantum gravity on the lattice this last fact
manifests itself in the non-closure of the algebra of constraints. 

The introduction of the Ashtekar new variables, in terms of which
canonical general relativity resembles a Yang-Mills theory revived the
interest in studying a lattice formulation of the theory 
\cite{ReSm,Re,Lo}. 
The new variables allow for a much cleaner
formulation, quite resemblant to that of Kogut and Susskind of gauge
theories.  However, it does little to attack the fundamental problem
we mentioned in the first paragraph. The lattice formulation still has
the problem of the closure of the constraints. The new variable
formulation has also allowed to obtain several formal results in the
continuum connecting knot theory with the space of solutions of the
Wheeler-DeWitt equation in loop space. This adds an extra motivation
for a lattice formulation, since in the lattice context these results
could be rigorously checked and many of the regularization ambiguities
present in the continuum could be settled.

It is not surprising that the lattice counterparts of diffeomorphisms do
not close an algebra.  A lattice diffeomorphism is a {\em finite}
transformation.  Finite transformations do not structure themselves
naturally into algebras but into groups.  One cannot introduce an
arbitrary parameter that measures ``how much'' the diffeomorphism shifts
quantities along a vector field.  One is only allowed to move things in
the fixed amounts permitted by the lattice spacing.  It is only in the
limit of zero lattice spacing, where the constraints represent
infinitesimal generators, that one can recover an algebra structure. 
The question is: does that algebra structure correspond to the classical
constraint algebra of general relativity? In general it will not. 

In this paper we present a lattice formulation of quantum gravity in the
loop representation constructed in such a way that from the beginning we
have good hopes that the diffeomorphism algebra will close in the limit
when the lattice spacing goes to zero.  The strategy will be the
following: in the continuum theory the solutions to the diffeomorphism
constraint in the loop representation are given by knot invariants.  We
will introduce a set of constraints in the lattice such that their
solution space is a lattice generalization of the notion of knot
invariants.  This means that the solutions, in the limit when the
lattice spacing goes to zero, are usual knot invariants.  As a
consequence the constraints are forced to become the continuum
diffeomorphism constraints and close the appropriate algebra.  Roughly
speaking, if this were not the case they could not have the same
solution space as the usual continuum constraints.  We will check
explicitly the closure in the continuum limit. Although this has already 
been achieved for other 
proposals of diffeomorphism contraints on the lattice \cite{Re} the main
advantage of our proposal is that our constraints not only close
the appropriate algebra in the continuum limit but also 
admit as solutions objects that reduce
in that  limit to usual knot invariants, including
the invariants from Chern--Simons theory

We will also introduce a Hamiltonian constraint in the space of
lattice loops and show that it has the correct continuum limit. It
remains to be shown if it satisfies the correct algebra relations with
the diffeomorphism constraint in the continuum limit.

The main purpose of this paper, however, is to explore how to implement
the diffeomorphism symmetry and the idea of knot invariants in the
lattice and its relationship to quantum gravity, both at a kinematical
and dynamical level.  We will consider explicit definitions for knot
invariants and polynomials, that are the lattice counterpart of formal
states of quantum gravity in the continuum formulation.  In particular
we will introduce the idea of linking number, self-linking number and
invariants associated with the Alexander-Conway polynomial.  We will
also discuss the construction in the lattice of a polynomial related to
the Kauffman bracket of the continuum and analyze the issue of the
transform into the loop representation of the Chern-Simons state and its
relation with the framing ambiguity.  We will notice that regular
isotopic invariants (invariants of framed loops) are not annihilated by
the diffeomorphism constraint in the lattice.  These results allow to
put on a rigorous setting many results that were only formally available
in the continuum. 

Moreover, we will introduce a Hamiltonian constraint in the lattice
whose action on knot invariants can be characterized as a set of skein
relations in knot space. This allows us to show rigorously that lattice
knot invariants are annihilated by the Hamiltonian constraint, again
confirming formal results in the continuum. We also point out
differences in the details with the continuum results.

Our approach departs in an important way from previous lattice
attempts \cite{ReSm,Re,Lo}: we will work directly in the loop
representation. There is a good reason for doing this. When one
discretizes a theory there are many ambiguities that have to be
faced. There are many discretized version of a given continuum
theory. If one's objective is to construct a loop representation it is
better to discretize at the level of loops than to try to discretize a
theory in terms of connections and then build a loop
representation. We will also start from a non-diffeomorphism
formulation at a kinematical level, since it is only in this context
that one can study the action of the Hamiltonian constraint of quantum
gravity, which is not diffeomorphism invariant. Moreover it is the
only framework in which one can address the issue of regular isotopy
invariants, which, strictly speaking, are not diffeomorphism
invariant. We will then go to a representation that is based on
diffeomorphism invariant functions and capture the action of the
Hamiltonian constraint and interpret it as skein relations in that
space. We end with a discussion of further possibilities of this
approach.

\section{Lattice loop representation: the diffeomorphism constraint}

\subsection{The group of loops on the lattice}

Consider a three dimensional manifold with a given topology, say $S^3$
and a coordinate patch covering a local section of it. We set up a
cubic lattice in this coordinate system (this is for simplicity only,
none of the arguments we will give depends on the lattice being
square, only on having the same topology as a square lattice).
The position of the sites are labeled with latin
letters $(m,\,n,...)$ where $m=(m_1,m_2,m_3)$ with $m_i$ integers, and
the links emerging from a given site are labeled by $u_{\mu}$ with
$\mu=(-3,-2, \ldots, 3)$, and $u_{\pm 1}=(\pm 1,0,0)$. The typical
coordinate lattice spacing will be denoted by $a$.

        Let us now introduce loops with origin $n_0$. A closed curve is
given by a finite chain of vectors

\begin{equation}
(u^1,....u^N)  \;\;\hbox{with}\;\; \sum_{i=1}^N u^i =0
\end{equation}
starting at $n_0$. In this paper we will use the word loop in a precise
way, analogous to the one that some authors use in the continuum. This
notion refers to the fact that there is more information in a closed
curve than that needed to compute the holonomy of a connection around
the curve. Therefore there are several closed curves that yield the same
holonomy. The equivalence class of such curves is to what we will refer
to as loops. In order to define loops on the lattice, 
we now introduce a reduction
process. We define $R(u^1,....u^N)$ as the chain obtained
by elimination of opposite successive vectors
\begin{equation}
(...u^i_\alpha,u^{i+1}_\mu,u^{i+2}_{-\mu},u^{i+3}_\beta...) \rightarrow
(...u^i_\alpha,u^{i+1}_\beta...).
\end{equation}

Once a couple of vectors is removed,new collinear opposite vectors may
appear and must be eliminated. The process is repeated until one gets an
irreducible chain. One can show that the reduction process is
independent of the order in which the vectors are removed. A loop
$\gamma$ is an irreducible closed chain of vectors starting at $n_0$.
There is a natural product law in loop space. The product of $\gamma_1$
and $\gamma_2$ is the reduced composition of their irreducible chains
and one can show that loops form a group\cite{GaTr81}. The inverse of
$\gamma_1$ is $\bar \gamma_1=(-u^N,....-u^1)$. We will consider a quantum 
representation of gravity in which wavefunctions are functions of the 
group of loops on the lattice $\Psi(\gamma)$. 

Loop representations require considering loops with intersections.
Therefore one is interested in intersecting loops on the lattice.  We
define the {\em intersection class} of a loop by assigning a number to
each intersection and making a list obtained by traversing the loop
and listing the numbers of the successive intersections traversed. Two
loops will belong to the same intersection class if the lists of
numbers are the same. The concept of intersection class will help
define later the transformations that behave like diffeomorphisms on
the lattice. Intersecting loops behave as a ``boundary'' between
different knot classes of non-intersecting loops.  We will therefore
require that the transformations do not ``cross the boundary'' by
requesting that they do not change the intersection class of the
loops. Any deformation of the loop that does not make it cut itself
preserves the intersection class. Wavefunctions in the loop
representation are cyclic functions of loops and therefore we will
identify loops belonging to the same intersection class through cyclic
rearrangements.

\subsection{The diffeomorphism constraint: continuum limit}

We now proceed to write a generator of deformations on the lattice,
which in the continuum limit will yield the diffeomorphism constraint.
We define an operator $d_\mu(n)$ that deforms the loop at the point
$n$ of the lattice as shown in figure 1. For example, the action of
the operator on a chain,
\begin{equation}
\gamma=(...u^i(n),u^{i+1}(n)...u^j(n),u^{j+1}(n)...u^k(n),u^{k+1}(n)...)
\end{equation}
is,
\begin{eqnarray}
\gamma_D \equiv d_\mu(n) \gamma \equiv
R(...u_\mu,u^i(n),u^{i+1}(n),u_{-\mu}...u_\mu,u^j(n),u^{j+1}(n),u_{-\mu} \\
...u_\mu,u^k(n),u^{k+1}(n),u_{-\mu}...). \nonumber
\end{eqnarray}
\begin{figure}
\hskip 4cm \psfig{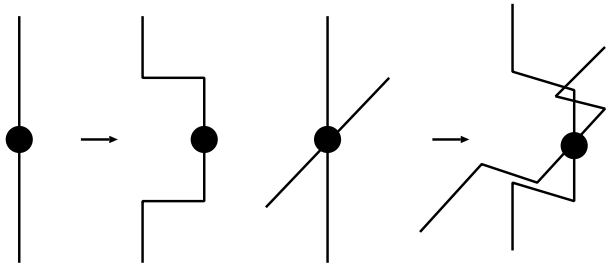}
\caption{The action of the deformation operator on a lattice loop at a
regular point and at an intersection.}
\end{figure}

In the above example we have a loop that goes three times through the
same point. If the point $n$ does not lie on the loop considered the 
deformation operator leaves the loop invariant. The action of the
deformation operator is to add two plaquettes one before and one after
the point at which it acts along the line of the loop. If the action
takes place at an intersection, as in the example shown, the
deformation adds two plaquettes along each line going through the
intersection.

We now define an {\em admissible deformation} as a deformation that
does not change the intersection type of the loop. Admissible
deformations are constructed with the operator $d\mu(n)$ but defining
its action to be unity in the case in which the loop would change its
intersection class as a consequence of the deformation. Typical
examples of deformations that change the intersection class arise when
one has parallel lines separated by one lattice spacing in the loop
and one deforms in a transverse direction. These deformations are set
to unity.

We now consider a quantum representation formed by wavefunctions
$\Psi(\gamma)$ of loops on the lattice. One can therefore introduce an
operator which implements the 
admissible deformations of the loops ${\cal D}_\mu(n) \Psi(\gamma)=
\Psi(d_\mu(n) \gamma)$ and an associated operator,

\begin{equation}
{\cal C}_\mu(n) \Psi(\gamma) 
\equiv {(1+P_\mu(n)) \over 4} \left({\cal D}_\mu(n) \Psi( \gamma)
-{\cal D}_{-\mu}(n) \Psi(\gamma)\right). 
\end{equation}
where the weight factor $P_\mu(n)$ counts the number of non-admissible
deformations that the displacements would create. If both displacements
are admissible the overall factor is $1/4$, if one is non-admissible, it
is $1/2$. This factor is needed to ensure consistency of the algebra of
constraints.

We will show that this operator, when acting on a holonomy on the
lattice, produces the usual diffeomorphism constraint of the Ashtekar
formulation in the continuum limit. We will take the continuum limit
in a precise way: we will assume that loops are left of a fixed length
and the lattice is refined. As a consequence of this, loops will never
have two parallel sections separated by only one plaquette. Due to
this, for any given calculation we only need to consider three
different kinds of points in the loop: regular points, corners and
intersections. The intersections can have corners or go ``straight
through''. 

Before going into the explicit computations it is worthwhile analyzing
up to what extent can one recover diffeomorphisms from a lattice
construction. After all, all deformations on the lattice are discrete
and are therefore {\em homeomorphisms}. It is clear that at regular
points of the loop there is no problem, the addition of plaquettes
becomes in the limit the infinitesimal generator of deformations, as
we shall see. The situation is more complicated at
intersections. Deformations that do not occur at the intersection, but
at points adjacent to it can change the nature of the
intersection. For instance, an intersection that goes ``straight
through'' can be made to have a kink by deforming at an adjacent
point. In the continuum, a diffeomorphism {\em cannot} change straight
lines into lines with kinks, therefore the above kind of deformations
must be forbidden if one wants the lattice transformations to
correspond to diffeomorphisms in the continuum. Because of the way we
are taking the continuum limit this situation does not occur, since
all points of the loop are either {\em at} an intersection or {\em far
away} from it. We will see, however, that this problem resurfaces when
one wants to compute the diffeomorphism algebra, and we will discuss
it there.

In order to see that the above operator corresponds in the continuum
to the generator of diffeomorphism we consider a holonomy along a
lattice loop $T^0(\gamma)\equiv {\rm Tr}[\prod_{l \in \gamma}U(l)]$ where
$U(l)=\exp{a A_{b}(n)}$, $a$ is the lattice spacing and $A_b(n)$ is
the Ashtekar connection at the site n. Then,
\begin{eqnarray}
{\cal C}_{\mu_0}(n) T^0(\gamma) = {(1+P_\mu(n))\over 4} 
[T^0(d_\mu \gamma)-T^0(d_{-\mu}\gamma)].
\end{eqnarray}

In the limit $a \rightarrow 0$ while the loop remains finite, this
action is always a local deformation (in the sense that there are no
lines of the loop in a neighborhood of each other). Assuming the
connections are smooth, we get,
\begin{eqnarray}
{\cal C}_{\mu}(n) T^0(\gamma) ={\textstyle {1\over 2}}\left( 
a^2 N_n(\gamma)Tr[F_{ab}(n)U(\gamma)]u_{\mu}^a u^b_{\nu}(n)\right.\\
\left.
+a^2N_n(\gamma)Tr[F_{ab}(n)U(\gamma)]
u_{\mu}^a u^b_{\nu'}(n)\right) \label{difinf} \nonumber
\end{eqnarray}
where $\nu$ and $\nu'$ represent the links adjacent to the site $n$ on
the loop and $N_n(\gamma)$ is a function that is $1$ if $n$ is on the loop
$\gamma$ and zero otherwise, so $N_n(\gamma)=\sum_{n'\in\gamma}
\delta_{n,n'}$ where $\delta_{n,n'}$ is a Kronecker delta.

But
\begin{equation}
\lim_{a \rightarrow 0} {{N_n(\gamma) a (u^a_\nu+u^a_{\nu'})}\over{2 a^3}}=
\int_\gamma dy^a\delta(x-y) \equiv X^a(x,\gamma) \label{tan}
\end{equation}
where $x=\lim_{a\rightarrow 0}  na$, $x=\lim_{a\rightarrow 0}  n'a$
and we have used that
$lim_{a\rightarrow 0} \delta_{na,n'a}/a^3 =\delta(x-y)$. Therefore
\begin{eqnarray}
\lim_{a \rightarrow 0}  \textstyle{1/a^4}{\cal C}_{\mu}(n)T^0(\gamma)=
u^a_{\mu}X^b(x,\gamma)\Delta_{ab}(\gamma^x)T^0(\gamma)=\\
u^a_{\mu} \int_\gamma dy^b \delta(x-y) \Delta_{ab}(\gamma^y)T^0(\gamma)=
u^a_{\mu} F_{ab}^i(x) {\delta \over \delta A_b^i} T^0(\gamma) =
u^a_\mu \hat{\cal C}_a T^0(\gamma)
\end{eqnarray}
which is the explicit form of the vector constraint 
$\hat{\cal C}_b = \hat{\tilde{E}}^a_i \hat{F}_{ab}^i$ 
in the Ashtekar formulation. 

This procedure is valid for any regular point of the loop or corner.
A similar procedure may be followed for a site including intersections
or corners. It can be straightforwardly verified that the vector
constraint is also recovered in those cases.

\subsection{The diffeomorphism constraint: constraint algebra}

In order to have a consistent quantum theory, one has to show that the
quantum constraint  algebra reproduces to leading order in $\hbar$ the
classical one. In a lattice theory, the objective is to show that the
quantum constraint algebra reproduces the classical one {\em in the
continuum limit}. Specifically, for the case of diffeomorphisms,

\begin{equation}
[\lim_{a\rightarrow 0} {{\cal C}_\mu(n)\over a^4},\lim_{a\rightarrow 0}
 {{\cal C}_\nu(n')\over a^4}] = \lim_{a\rightarrow 0} {1\over a^8}
[{\cal C}_\mu(n),{\cal C}_\nu(n')].
\end{equation}

We will here show that the correct algebra is reproduced in the limit
in which the lattice spacing goes to zero. This is an important calculation,
since it is not obvious that diffeomorphism symmetry can be implemented in
a square lattice framework as we propose here. We will see that there are
subtle points in the calculation. We will present the explicit calculation
for a regular point of the loop only in an explicit fashion. Even for that
case the calculation is quite involved. 
\begin{figure}[t]
\hskip 1cm \psfig{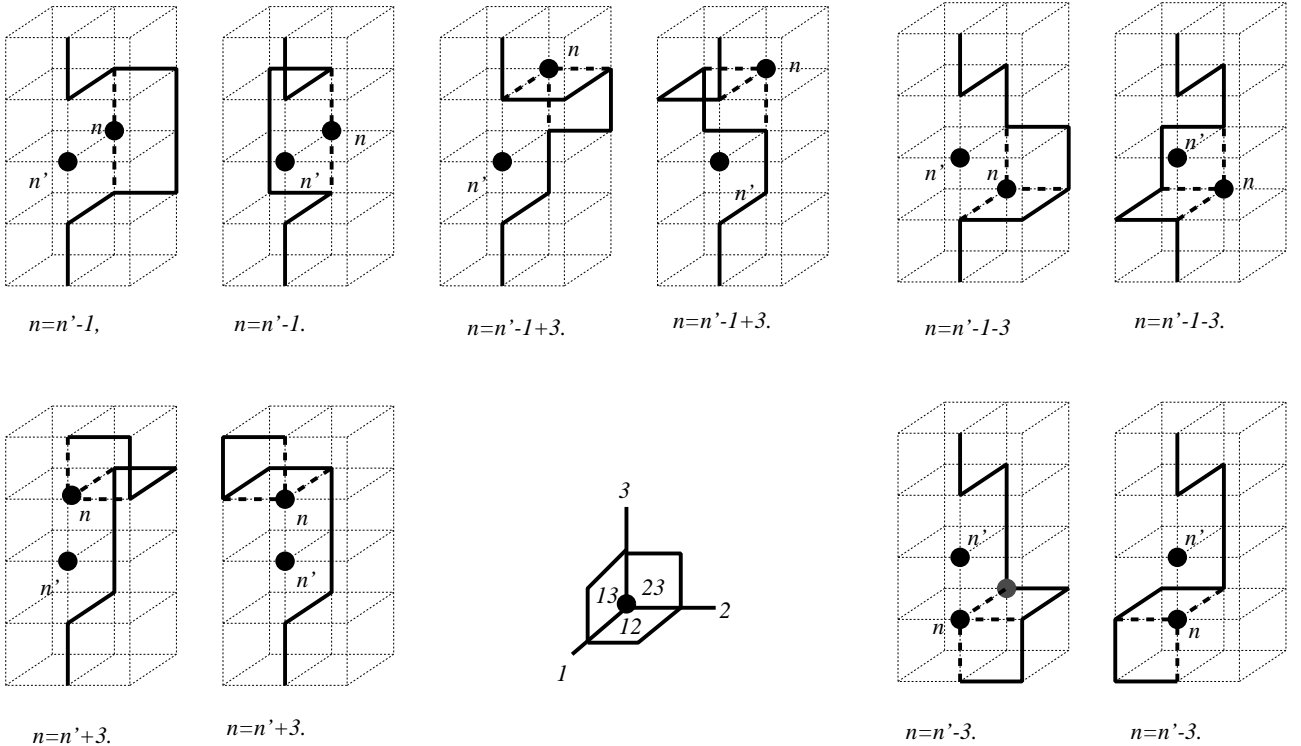}
\caption{The action of the first portion of terms in the commutator of 
two diffeomorphism constraints acting at a regular lattice point.}
\label{deformaprimero}
\end{figure}

Let us now consider the algebra of the operators ${\cal C}_\mu(n)$ and
${\cal C}_\nu(n')$ and its continuum limit. As before, we assume that
a refinement of the lattice has taken place with the loop length
remaining finite in such a way that any point of the loop is either in
a corner, intersection or regular point. We will only discuss
explicitly the case in which the commutator is evaluated at a regular
point. To simplify the notation, we will consider $\mu$ and $\nu$ in
directions perpendicular to the loop and we take three coordinate axes
$1,2,3$ with $3$ parallel to the loop and the other two perpendicular.
We will compute $[\hat{C}_2(n),\hat{C}_1(n')] \psi(\gamma)$. Let us
consider first the term $\hat{C}_2(n)\hat{C}_1(n')$. In order for it
to be nonvanishing, the point $n'$ has to be on the loop $\gamma$. 
The first diffeomorphism generates two
terms, corresponding to the addition of two plaquettes in the forward
$1$ direction and backward. The
second diffeomorphism will
lead to non-zero commutator 
if the point $n$ lies in
one of the points marked in figure \ref{deformaprimero} on the
deformed loop. There are five possible such points along the
deformation.  At each the action of the second diffeomorphism
generates two terms.  The resulting ten deformed loops are shown
explicitly in the figure \ref{deformaprimero}. There will be ten similar
terms resulting from the ``backwards'' action of the first
diffeomorphism. To clarify the calculation, let us concentrate on the
first pair of terms displayed in figure \ref{deformaprimero}. The
infinitesimal deformation of the original loop $\gamma$ generated
by the loop derivative operator \cite{Gaplb91}
is represented through the introduction in the holonomy of 
field strengths $F_{ab}(P,p)$
, depending on a path $P$ and its end point $p$, 
contracted with the element of area of the plaquette. 
To unclutter the notation, 
when a plaquette in the direction $1-2$ is added, we will
denote this as $F_{12}(p)$ dropping the dependence in the path $P$ 
(we assume the path goes from the basepoint of the loop to the point 
of interest). 
With this notation, the contribution of the first two 
terms of figure \ref{deformaprimero} is (we are neglecting
all the contributions of powers greater than $a^2$),
\begin{eqnarray}
N_\gamma(n') &\{& {\rm Tr}[\, (\, F_{23}(n'-1)+F_{23}(n'-1)
\,) U(\gamma) ] \nonumber\\&&-
{\rm Tr}[ \, (-F_{23}(n'-1)-F_{23}(n'-1) \, ) U(\gamma) ]\}
\end{eqnarray}
where by $-1$, $-3$, etc we denote the point displaced one lattice
unit in the corresponding direction, the sign indicating forward
or backward respect to the orientation chose in the trihedron shown in
the figure.

We see that the contribution of the plaquettes added by the first 
diffeomorphism, namely the $F_{13}'s$ cancel each other at 
this order. Taking into
account these cancellations, the result of all the terms considered
in figure \ref{deformaprimero} is,
\begin{eqnarray}
N_\gamma(n') &\{&\delta(n-n'+1) 
{\rm Tr}[ (F_{23}(n'-1)+F_{23}(n'-1)) U(\gamma)]\nonumber\\
&&-\delta(n-n'+1) 
{\rm Tr}[ (-F_{23}(n'-1)-F_{23}(n'-1)) U(\gamma)]\nonumber\\
&&+\delta(n-n'+1-3) 
{\rm Tr}[ (F_{23}(n'-1+3)-F_{12}(n'-1+3)) U(\gamma)]\nonumber\\
&&-\delta(n-n'+1-3) 
{\rm Tr}[ (-F_{23}(n'-1+3)+F_{12}(n'-1+3)) U(\gamma)]\nonumber\\
&&+\delta(n-n'+1+3) 
{\rm Tr}[ (F_{12}(n'-1-3)+F_{23}(n'-1-3)) U(\gamma)]\nonumber\\
&&-\delta(n-n'+1+3) 
{\rm Tr}[(-F_{12}(n'-1-3)-F_{23}(n'-1-3)) U(\gamma)]\nonumber\\
&&+\delta(n-n'-3) 
{\rm Tr}[(-F_{12}(n'+3)+F_{23}(n'+3)) U(\gamma)]\nonumber\\
&&-\delta(n-n'-3) 
{\rm Tr}[(F_{12}(n'+3)-F_{23}(n'+3)) U(\gamma)]\nonumber\\
&&+\delta(n-n'+3) 
{\rm Tr}[(F_{23}(n'-3)+F_{12}(n'-3)) U(\gamma)]\nonumber\\
&&-\delta(n-n'+3) 
{\rm Tr}[(-F_{23}(n'-3)-F_{12}(n'-3)) U(\gamma)]\}
\label{eq:D_-1}
\end{eqnarray}
where we have, to make more direct the continuum analysis used the
notation of Dirac deltas for what strictly speaking are Kronecker deltas
at this stage of the calculation.

\begin{figure}[t]
\hskip 1cm \psfig{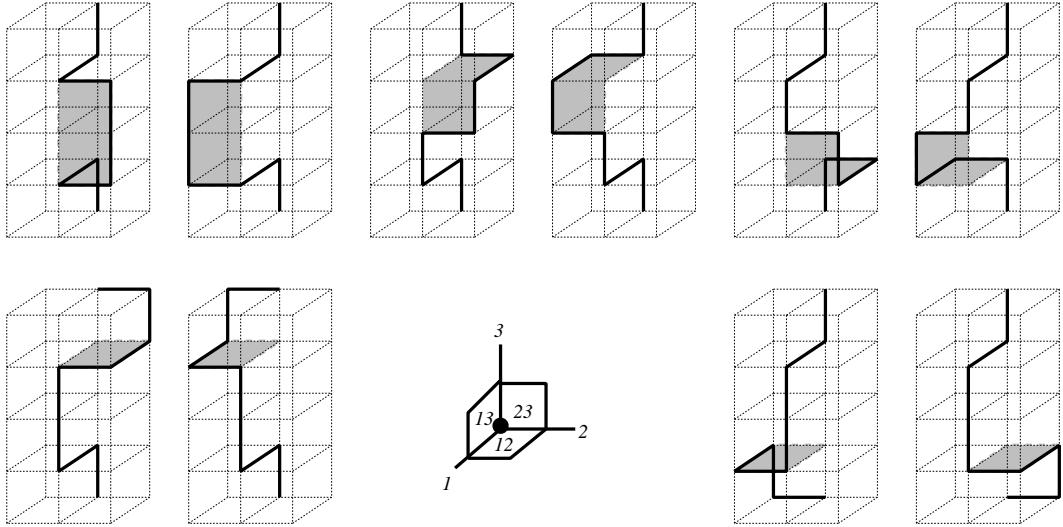}
\caption{The second portion of the terms of the commutator of two 
diffeomorphisms at a regular point of the lattice.}
\label{deforsegundo}
\end{figure}

We now need to consider the terms from the ``backwards'' action of the
first diffeomorphism constraint. This gives rise to the contributions
depicted in figure \ref{deforsegundo} and are explicitly given by,
\begin{eqnarray}
N_\gamma(n') &\{&\delta(n-n'-1) 
{\rm Tr}[ (F_{23}(n'+1)+F_{23}(n'+1)) U(\gamma)]\nonumber\\
&&-\delta(n-n'-1) 
{\rm Tr}[ (-F_{23}(n'+1)-F_{23}(n'+1)) U(\gamma)]\nonumber\\
&&+\delta(n-n'-1-3) 
{\rm Tr}[ (F_{23}(n'+1+3)+F_{12}(n'+1+3)) U(\gamma)]\nonumber\\
&&-\delta(n-n'-1-3) 
{\rm Tr}[ (-F_{23}(n'+1+3)-F_{12}(n'+1+3)) U(\gamma)]\nonumber\\
&&+\delta(n-n'-1+3) 
{\rm Tr}[ (-F_{12}(n'+1-3)+F_{23}(n'+1-3)) U(\gamma)]\nonumber\\
&&-\delta(n-n'-1+3) 
{\rm Tr}[(F_{12}(n'+1-3)-F_{23}(n'+1-3)) U(\gamma)]\nonumber\\
&&+\delta(n-n'-3) 
{\rm Tr}[(F_{12}(n'+3)+F_{23}(n'+3)) U(\gamma)]\nonumber\\
&&-\delta(n-n'-3) 
{\rm Tr}[(-F_{12}(n'+3)-F_{23}(n'+3-2)) U(\gamma)]\nonumber\\
&&+\delta(n-n'+3) 
{\rm Tr}[(F_{23}(n'-3)-F_{12}(n'-3)) U(\gamma)]\nonumber\\
&&-\delta(n-n'+3) 
{\rm Tr}[(-F_{23}(n'-3)+F_{12}(n'-3)) U(\gamma)]\}
\label{eq:D_+1}
\end{eqnarray}

The terms in the above two expressions can be combined to form
a series of differences of delta's times $F$'s, 
\begin{eqnarray}
4N_\gamma(n') &\{& 
\Delta_1 [ \;\; 2\delta(n'-n) {\rm Tr}[F_{23}(n'))U(\gamma)]\nonumber\\ 
&&+\delta(n'-n-3) {\rm Tr}[(F_{23}(n'+3)U(\gamma)]\nonumber\\ 
&&+\delta(n'-n+3) {\rm Tr}[F_{23}(n'-3)U(\gamma)] \;\;]\nonumber\\ 
&&+\Delta_3 [2\delta(n'-n) {\rm Tr}[F_{12}(n')U(\gamma)]\nonumber\\
&&+\delta(n'-n-1) {\rm Tr}[F_{12}(n'+1)U(\gamma)]\nonumber\\
&&+\delta(n'-n+1) {\rm Tr}[F_{12}(n'-1)U(\gamma)] \;\; ] \},
\label{eq:half-conmutator}
\end{eqnarray}
where the differences $\Delta_\mu$ mean evaluate the quantity at 
$n'$ and $n'+\mu$ and take the difference.
The derivatives of $F_{23}$ along the direction 1 are formed by the
first six terms of (\ref{eq:D_+1}) and (\ref{eq:D_-1}) 
(the term multiplied by 2 comes from the first two lines
and the remaining from the next four lines). 
The last four terms give vanishing 
contributions for $F_{23}$. The derivatives along $3$ of $F_{12}$ 
are produced in two identical copies 
%in an analogous way as that of the 
%the derivatives of $F_{23}$ along the direction 1.
by (\ref{eq:D_+1}) and (\ref{eq:D_-1}), combining the
contribution of each line with the one two lines below in $F_{12}$.

To understand this result it is useful to pay attention to the shading
in figure \ref{deforsegundo}, since we see that the contributions to
the action of the successive operators is mirrored by the shaded areas
of the figure. All the unshaded plaquettes cancel each other in the 
various terms. Only the shaded ones survive. For instance, the $2-3$ 
shaded plaquettes to the front, of which we have two in the first
drawing of figure \ref{deforsegundo}, two in the second and one in
the third, fourth, fifth and sixth, minus the corresponding terms of
figure \ref{deformaprimero} give rise to the first, second and third
terms of equation (\ref{eq:half-conmutator}). 

Equation (\ref{eq:half-conmutator})
can be rewritten as
\begin{eqnarray}
4N_\gamma (n') &\{& \Delta_1 [ \; 
2\delta(n-n'){\rm Tr}[F_{23}(n')U(\gamma)]  \nonumber \\
&&+ \delta(n-n'-3){\rm Tr}[F_{23}(n'+3)U(\gamma)] \nonumber \\
&&+\delta(n-n'+3){\rm Tr}[F_{23}(n'-3)U(\gamma)] \; ] \nonumber \\
&&+ \Delta_3 [ \; 2\delta(n-n'){\rm Tr}[F_{12}(n')U(\gamma)]  \nonumber \\
&&+ \delta(n-n'-1){\rm Tr}[F_{12}(n'+1)U(\gamma)] \nonumber \\
&&+\delta(n-n'+1){\rm Tr}[F_{23}(n'-1)U(\gamma)] \; ] \, \}.
\label{eq:half-conm2}
\end{eqnarray}
In order to write the commutator
$[\hat{C}_2(n),\hat{C}_1(n')] \psi(\gamma)$
we have to subtract to (\ref{eq:half-conm2})
the same expression but interchanging
$1\leftrightarrow 2$ and 
$n \leftrightarrow n'$. If we now express the $N_\gamma(p)$
as $\sum_{m\in \gamma} \delta(p-m)$ then we get
\begin{eqnarray}
[\hat{C}_2(n),\hat{C}_1(n')] =
{1\over 4}\sum_{m' \in \gamma} \delta(n'-m')
\{ \Delta_1 [ \; 2\delta(n-m'){\rm Tr}[F_{23}(m')U(\gamma)] \nonumber \\
+ \delta(n-m'-3){\rm Tr}[F_{23}(m'+3)U(\gamma)] \nonumber \\
+\delta(n-m'+3){\rm Tr}[F_{23}(m'-3)U(\gamma)] \; ] \nonumber \\
+ \Delta_3 [ \; 2\delta(n-m'){\rm Tr}[F_{12}(m')U(\gamma)]  \nonumber \\
+ \delta(n-m'-1){\rm Tr}[F_{12}(m'+1)U(\gamma)] \nonumber \\
+\delta(n-m'+1){\rm Tr}[F_{23}(m'-1)U(\gamma)] \; ] \, \} \nonumber \\
-{1\over 4}\sum_{m \in \gamma} \delta(n-m) \{ \Delta_1 [ \;
2\delta(n'-m){\rm Tr}[F_{13}(m)U(\gamma)] \nonumber \\ 
+ \delta(n'-m-3){\rm Tr}[F_{13}(m+3)U(\gamma)] \nonumber \\
+\delta(n'-m+3){\rm Tr}[F_{13}(m-3)U(\gamma)] \; ] \nonumber \\
+ \Delta_3 [ \; 2\delta(n'-m){\rm Tr}[F_{21}(m)U(\gamma)] \nonumber \\
+ \delta(n'-m-2){\rm Tr}[F_{21}(m+2)U(\gamma)] \nonumber \\
+\delta(n'-m+2){\rm Tr}[F_{21}(m-2)U(\gamma)] \; ] \, \}
\label{eq:conmutator}
\end{eqnarray}

If we now neglect terms of higher order in the lattice spacing, we can
move all the delta's and $F$'s to the same locations and, taking into
account the explicit dependence on the lattice spacing $a$
the equation (\ref{eq:conmutator}) becomes
\begin{eqnarray}
[\hat{C}_2(x),\hat{C}_1(y)] = 
\lim_{a\rightarrow 0} [{\hat{C}_2(na)\over a^4},{\hat{C}_1(n'a) \over a^4}]
=  \lim_{a\rightarrow 0}  {1\over a^6} \sum_{m \in \gamma} \nonumber \\
&&\{ \; \;  \delta(n'-m)
\Delta_1  \delta(n-m)  {\rm Tr}[F_{23}(m)U(\gamma)]
\nonumber \\
&&- \Delta_2 \delta(n'-m) \delta(n-m) {\rm Tr}[F_{13}(m)U(\gamma)] 
\nonumber \\
&&+ \delta(n'-m) \delta(n-m)
\Delta_1  {\rm Tr}[F_{23}(m)U(\gamma)]  \nonumber  \\ &&
+ \delta(n'-m) \delta(n-m)
\Delta_2  {\rm Tr}[F_{31}(m)U(\gamma)]  \nonumber \\ &&
+ \delta(n'-m) \delta(n-m)
\Delta_3  {\rm Tr}[F_{12}(m)U(\gamma)]  \nonumber \\ &&
+\Delta_3  \delta(n'-m)  \delta(n-m)
{\rm Tr}[F_{12}(m)U(\gamma)] \nonumber \\ &&
+ \delta(n-m) \Delta_3  \delta(n'-m)
{\rm Tr}[F_{12}(m)U(\gamma)] \nonumber \\ &&
+ \delta(n-m) \delta(n'-m) \Delta_3
{\rm Tr}[F_{12}(m)U(\gamma)] \; \}
\label{eq:conmutator2}
\end{eqnarray}
where the power $1/a^6$ arises from the fact that implicit in all
the previous calculations was an $a^2$ power, as we mentioned at the 
beginning.
The third, fourth and fifth terms of (\ref{eq:conmutator2})
cancel out by virtue of the Bianchi identity. The 
sixth, seventh and eighth form a total derivative
along the 3-direction.

Recalling that $\delta(na-ma)/a^3 \rightarrow \delta(x-z)$ and
$\Delta_1 \delta(na-ma)/a^4 \rightarrow \partial_1 \delta(x-z)$
we arrive to 
\begin{eqnarray}
[\hat{C}_2(x),\hat{C}_1(y)] = \nonumber \\
&&\{ \; \partial_1 \delta(y-x)
\int_\gamma dz \delta(y-z) {\rm Tr} [F_{23}(y)U(\gamma)] \nonumber \\
&&-\partial_2 (\, \delta(x-y) \,) \int_\gamma dz \delta (x-z)
{\rm Tr} [F_{13}(x)U(\gamma)] \; \}.
\end{eqnarray}

The calculation displayed above is true at a regular point of the
loop. A similar calculation can be performed at corners with the same
result. A problem arises, however, at intersections. This is related
to the issue we discussed before of homeomorphisms vs
diffeomorphisms. As we pointed out, the operator we introduced
generates diffeomorphisms at regular points of the loop or at
intersections or corners. A problem develops if one acts in points
that are immediately adjacent to intersections, since the deformation
can change the character of intersections (introducing kinks). One
cannot ignore this fact in the commutator algebra. When one acts with
two successive diffeomorphisms at an intersection, it is inevitable to
consider one of the operators as acting at a point adjacent to the
intersection {\em before taking the continuum limit}. This leads to
changes in the character of the intersection that we do not want to
allow. If we did so, we would not recover diffeomorphism in the
continuum but homeomorphisms, which can introduce kinks in the
intersection. If one allows these kind of deformations in the lattice,
the calculation of the algebra at intersections goes through with no
problem. In the continuum, the only difference between the
homeomorphism and diffeomorphism algebra is given by the nature of the
smearing functions of the constraints, so it is not surprising that we
recover the same algebra. If one restricts the
action of the deformations in order to avoid changing the character of
intersections, by defining the action of the operator to be unity if
it changes the character of the intersection, the commutator algebra
fails at intersections. More precisely, the constraints {\em commute}
at intersections. Because this failure of the constraint algebra
occurs only at a zero measure set of points, one can still consider
the quantum theory to be satisfactory, since one is smearing the
constraints with smooth functions and therefore cannot distinguish
this algebra from one with different values of the commutators at a
zero measure set of points. We will adopt this latter point of view in
this paper, since our primary motivation is to implement
diffeomorphism symmetry in the continuum. This will also have a
practical consequence: because homeomorphism invariance is more
restrictive than diffeomorphism invariance we will see that it is
considerably easier to find invariants under diffeomorphisms than under
homeomorphisms.

\section{Knots on the lattice}

In this section we will discuss certain ideas of knot theory, in
particular those which pertain to knot invariants derived from
Chern--Simons theory, in the context of the lattice regularization.  We
define a knot invariant on the lattice as a quantity dependent on a loop
that is invariant under admissible deformations of the loops.  

The motivation for all this is that several of the knot invariants that
arise from Chern--Simons theory have difficulties with their definition. 
These difficulties are generically known as framing ambiguities.  They
refer to the fact that certain invariants are not well defined for a
single loop or their definition might have difficulties when extended to
loops with intersections.  The usual solution to these difficulties is
to ``frame" the loops, ie, convert them to ribbons.  It is the
mathematical language version of the problem of regularization in
quantum field theory.  We will see that the lattice treatment provides a
natural framing for knot invariants.  We will show that the results
obtained are in line with results obtained in the continuum and we will
see that the framing provided excludes regular isotopic invariants from
being candidates for states of quantum gravity.  The framed invariants
will be well defined, but will not be invariant under diffeomorphisms on
the lattice.

\subsection{The framing problem in the continuum}

In order to consider concrete knot invariants on the lattice we will
give lattice analogues of constructions that lead to knot invariants in
the continuum.  Let us therefore start by briefly recalling some of the
ideas that lead to knot invariants in the continuum.  An important
development in the last years has been the realization that topological
field theories can be powerful tools to construct explicit expressions
for knot invariants.  In particular, it was shown by Witten that in a
gauge theory given by the Chern-Simons action,

\begin{equation}
S_{CS} = {\rm Tr}\left[ \int d^3x \epsilon^{abc} (A_a \partial_b A_c
+{\textstyle {2 \over 3}} A_a A_b A_c)\right]
\end{equation}
the expectation value of the Wilson loop $W_\gamma[A]= {\rm Tr} P
\exp\left(i \oint_\gamma dy^a  A_a\right)$ ,
\begin{equation}
<W(\gamma)> = \int DA W_\gamma[A] \exp\left(i {k\over 8\pi} S_{CS}\right) 
\end{equation}
is an expression that satisfies the skein relation of the regular
isotopic knot invariant known in the mathematical literature as the
Kauffman bracket, evaluated for a particular combination of its
variables related to $k$.

A simpler version of the above statement is the observation that for
an Abelian Chern-Simons theory $S_{CS}^{\rm Abel} = {\textstyle {k\over
8\pi}} 
\int d^3x
\epsilon^{abc} A_a \partial_b A_c$ the expectation value of 
a Wilson loop is the exponential of the 
Gauss (self) linking number of the loop,
\begin{equation}
\exp\left({\textstyle {i \over 2 k}} \oint_{\gamma} dx^a 
\oint_{\gamma}dy^b \epsilon_{abc} {(x-y)^c\over |x-y|^3}\right).
\end{equation}

This expression is reminiscent of the linking number of {\em two}
loops,
\begin{equation}
{\textstyle {1 \over 4\pi}} \oint_{\gamma_1} dx^a 
\oint_{\gamma_2} dy^b \epsilon_{abc} {(x-y)^c\over |x-y|^3},
\end{equation}
a quantity that measures the number of times one loop ``threads
through'' the other. There is however, an important difference: in the
latter expression the two integrals are along different loops and the
points $x$ and $y$ never coincide. In appearance, therefore, the
self-linking number is ill-defined. 
In spite of its appearance, the integral is finite, but is ambiguously
defined, its definition requires the introduction of a normal to the
loop, which is not a diffeomorphism invariant concept. This problem is
known as the framing ambiguity and it is clearly related to
regularization. Suppose one attempts to define the self-linking number
by considering the linking number of a loop as the linking number of
the loop with a ``copy'' of itself, obtained by displacing the loop an
infinitesimal amount. This clearly regularizes the integral. However,
the result depends on how one creates the ``copy'' of the loop and how
it winds around the original loop. One of the main purposes of this
paper will be to show how to address the framing ambiguity in the
lattice. We will start by discussing the Abelian case.

The above problem is more acute if the loops have intersections. In
that case the integral tha appears in the self-linking number is not
even finite. Again, one can in principle solve the problem through a
mechanism of framing. An important aspect of the lattice construction
will therefore be to provide a precise mechanism for framing knot
invariants with intersections. 

Why is one interested in these particular invariants?. There are two
reasons. On one hand, because they are constructed as loop transforms
of quantities in terms of connections these invariants automatically
satisfy the ``Mandelstam constraints'' that must be satisfied by
wavefunctions of quantum gravity in the loop representation. This
makes them candidates for quantum states of gravity. Moreover, it has
been concretely shown that some of the invariants that arise due to
Chern-Simons theory formally solve the Hamiltonian constraint of
quantum gravity in the continuum. The lattice framework is an
appropriate environment to discuss these solutions in a rigorous
setting. 

\subsection{Abelian Chern-Simons theory on the lattice}

Let us start with the simple case of a $U(1)$ Chern-Simons theory and
study its lattice version. Although this is not directly connected
with quantum gravity we will see that it already contains several
ingredients we are interested in and in particular it allows the
discussion of the self-linking number, which plays a central role in
quantum gravity in the continuum. 

The Chern-Simons invariant can be written naturally in the lattice in
terms of a wedge product, as discussed in references
\cite{FrMa,Po}. In order to give more details we need to briefly introduce
the differential form calculus on the lattice \cite{BeJo}. 

A $k$-form
on the lattice is a function associated with a $k$-dimensional cell
$C_k$ in
the lattice that is skew-symmetric with respect to the orientation of
the cell. For instance, a zero-form is a scalar field (associated with
lattice sites), a one-form is a variable associated with lattice links
that changes sign when one changes the orientation of the link. 
A two-form is a variable associated with a plaquette, with sign
determined by the orientation of the plaquette. The exterior
derivative is a map between $k$ and $k+1$ forms defined by,
\begin{equation}
d\phi(C_{k+1}) = \sum_{C'_{K} \in \partial C_{k+1}} \phi(C'_k)
\end{equation}
where $\phi(C_n)$ is an $n$-form and the sum is along the cells that
form the boundary of $C_{k+1}$. For instance, given a one form
$A(C_1)$, we can define a two-form $F(C_2)$ obtained by summing $A$
along all the links that encircle the plaquette $C_2$ and which we
denote as $F=d A$. 

One can also introduce a co-derivative operator $\delta$ which associates a
$k-1$ form with a $k$ form, through
\begin{equation}
\delta \phi(C_{k-1}) = \sum_{\forall C_k / C_{k-1}\in \partial C_k}
\phi(C_{k}).
\end{equation}

{}From the above definitions one can see that both operators are
nilpotent, 
\begin{equation}
\delta^2=0,\qquad\qquad d^2=0,
\end{equation}
and in terms of these operators one can define a Laplacian, which
maps $k$-forms to themselves,
\begin{equation}
\nabla^2 \equiv \delta d+d \delta.\label{defdelta}
\end{equation}

To conclude these mathematical preliminaries we introduce a notion of
inner product on the lattice. The inner product of two $k$-forms is
defined by,
\begin{equation}
<\phi,\varphi> = \sum_{\forall C_k} \phi(C_k) \varphi(C_k),
\end{equation}
and this product has the property of ``integration by parts'',
\begin{equation}
<d \phi,\varphi> = <\phi,\delta \varphi>.
\end{equation}

With these elements we are able to introduce the Chern-Simons form
through the definition of a wedge product that associates (in three
dimensions) a one-form to each two-form and vice-versa. Given a
two-form associated with a plaquette $C_2$, the wedge product defines
a one-form associated with a link $C_1$ orthogonal to $C_2$, whose
value is given by the value of the one-form evaluated on
$C_2$. Evidently there is an ambiguity in how to define the wedge
product since there are eight links orthogonal to each plaquette. We
will see that different definitions of the wedge product will
correspond to different framings in the context of knot
invariants. There exist several definitions which all imply the
following identities,
\begin{eqnarray}
\# d A &=& \delta \# A,\label{hash1}\\
\# \delta A &=& d \# A.\label{hash2}
\end{eqnarray}

One possible consistent way to define the $\#$ operation is to assign
to a two-form associated with an  oriented plaquette an average of 
the one-forms associated with each of the eight links perpendicular to
the plaquette along its perimeter. This differs from the definition
taken in \cite{Po} which assigns only one link, but we will adopt it
in this paper because it yields a more symmetric framing of the knot
invariants. 

With the above notation, the Chern-Simons form in the lattice can be
written as,
\begin{equation}
S^{Abel}_{CS} = i k <F, \# A>. 
\end{equation}
This action is gauge invariant under transformations $A \rightarrow
A+d \xi$, since the transformation implies 
\begin{equation}
S_{CS}\rightarrow S_{CS} +
i k <F,\#d\xi> = S_{CS} +
i k <F,\delta \#\xi> = S_{CS} +
i k <d F,\#\xi> = S_{CS} +i k <d^2 A,\#\xi> = S_{CS}.
\end{equation}

One can now proceed to compute the expectation value of a Wilson loop
in a Chern-Simons theory on the lattice,
\begin{equation}
<W(\gamma)> = \int DA \exp\left(S^{\rm Abel}_{CS}\right) T^0(\gamma),
\label{cslat}
\end{equation}
which, after a straightforward Gaussian integration yields,
\begin{equation}
\exp\left(-{i\over 2 k} <\# l_\gamma,d \nabla^{-2} l_\gamma>\right)
\end{equation}
where $l_\gamma(C_1)$ is a one-form such that it is one if $C_1$
belongs to the loop $\gamma$ and zero otherwise. This expression takes
integer values and has a simple interpretation. In order to see this,
recall that in $S^3$ (or any simply connected region of an arbitrary
manifold), the one form $l$ satisfies $\delta l=0$ and therefore
can be written as a co-derivative of a two form $m$,
\begin{equation}
l=\delta m\label{gradiente}
\end{equation}
and substituting in the expression (\ref{cslat}) and using
(\ref{hash1}) and (\ref{defdelta}) expression (\ref{cslat}) reduces
to,
\begin{equation}
<\# m_\gamma,l_\gamma>.
\end{equation}

This expression is the particularization to one loop of the linking
number on the lattice, which we discuss in detail in the next section
and is usually called the self-linking number.  We will also return, at
the end of it, to the self-linking number. 

\subsection{The Gauss linking number in the lattice and self-linkings}

The linking number of two loops $\eta$ and $\gamma$ on the lattice is
given by
\begin{equation}
N(\eta,\gamma) = <\# m_\eta,l_\gamma>.\label{latlink}
\end{equation}
To understand this expression better notice that the simplest solution to
equation (\ref{gradiente}) consists of a set of plaquettes (it is easy to
see that the result is independent of the set chosen) with boundary
coincident with $\eta$ and consider an $m$ which is equal to $1$ on
each plaquette and $0$ otherwise. It is immediate from the definition
of $\delta$ that $\delta m$ only has contribution on the boundary of
the loop. As a consequence, the one form $\# m$ takes value on all
links orthogonal to the plaquettes considered. Therefore the inner
product of these one-forms with $l_\gamma$ will be non-vanishing
only if $\gamma$ threads through $\eta$, which is the
traditional definition of the linking number of two loops. The result
of the calculation is $1$ if the loops thread once each other and
counts the number of threadings in the case of multiple threadings. If
the loops are not linked the result is zero.

Notice that the above definition also includes as a particular case
that of two intersecting loops. To simplify the calculation, notice
that in this case the contribution is equal to counting the number of
links perpendicular to one of the loops that belong to the other
loop and dividing by eight. For instance, in the case of two planar
loops intersecting in such a way that one is perpendicular to the
plane of the other, the contribution is $1/2$. If the same
intersection occurs at the corner of one of the loops, the
contribution is $1/4$. We therefore see that the expression for the
linking number on the lattice is automatically ``framed''. Recall that
the definition of the linking number in the continuum was ill-defined
for intersecting loops. The lattice provides a prescription to assign
values to the linking number in the case of intersections. Evidently,
the particular ``framing'' obtained depends on the definition for the
$\#$ product considered. If instead of summing over the eight links
associated with each plaquette and dividing by eight we had had chosen
just one link (as is done in reference \cite{Po}) the result would be
different. Such a choice implies a preferred orientation in the
lattice.  The linking number of two intersecting loops would depend
on which side of the loop with respect that preferred orientation. 

The invariant discussed above (with the particular framing given for
the intersections) is invariant under the type of deformations that we
considered as analogues of diffeomorphism in the lattice in section
2. Those deformations did not change the type of intersections of
loops. Therefore the value of the linking number for intersecting
loops is invariant. Had we admitted the homeomorphisms as the symmetry
of the theory instead, the linking number we defined above would not
have been invariant for intersecting loops, since its value depends on
the particular type of intersection considered. 

Contrary to what happens in the continuum, in the lattice the
self-linking number defined as the linking number of a loop with itself
is a well defined quantity without the introduction of any external
framing.  If the loop is planar, the self-linking number vanishes.  If
not, it will be determined by the kinks and self-intersections of the
loop.  The evaluation in practice of the self-linking number can be done
just applying the definition of the linking number twice with the same
loop.  However the resulting quantity is {\em not} invariant under the
diffeomorphisms on the lattice we have considered.  To see this,
consider a planar loop and apply a deformation perpendicular to the
plane of the loop.  Due to definition (\ref{latlink}) the only
nonvanishing contributions come from the two links emerging
perpendicular to the loop in the deformation.  At regular points, each
link has a contribution equal and opposite in sign and cancel each
other.  However, at corners, the link at the corner contributes $1/2$ of
what the other does (in the definition of the $\#$ product introduced
corners of a loop only have one associated plaquette instead of two as a
regular point) and the self-linking number is therefore not invariant. 

There are several attitudes possible in face of this non-invariance of
the self-linking number. One could try to limit the action of
diffeomorphism such that they do not deform corners. This is
unacceptable. Corners are generic points in the lattice and
diffeomorphisms must act at them. Consider a large square loop in the
continuum. If one considers its lattice representation and 
aligns it with the lattice directions, it has only
four corners. However, if one rotates it a bit, many new corners are
introduced. Therefore diffeomorphisms must act at corners. Another
possibility would be to alter the definition of the $\#$ operation
in such a way as to treat corners in the same footing as regular
points. Although we cannot rule out such a re-definition, we have been
unable to find a suitable one. 

We therefore see that the lattice is useful to solve the problem of
framing of intersections for invariants of intersecting loops. It is
however, unable to solve the framing problem of regular isotopic
invariants. The corresponding quantities in the lattice are simply not
invariant. 

One last point of attention should be that we have not proven that the
Chern-Simons form that we introduced in the lattice is invariant under
the diffeomorphisms we considered. In order to do that we would need a
definition of the constraint in the connection representation, which
we do not have. It could be possible that the Chern-Simons form
considered is not invariant under the discrete diffeomorphisms 
and there lies the root of the non-invariance of the self-linking number. 

Finally, it is worthwhile pointing out that since the linking number
is obtained from an Abelian theory it displays a certain set of
``Abelian-like'' properties in terms of loops. For instance,
with respect to composition of curves and retracings,
\begin{eqnarray}
N(\gamma_1\circ\gamma_2,\gamma_3)&=&N(\gamma_1,\gamma_3)+
N(\gamma_2\circ\gamma_3) \label{abel1}\\
N(\gamma,\eta^{-1})&=&-N(\gamma,\eta),\label{abel2}
\end{eqnarray}
these properties will be crucial for the results we will derive in the
next section.

In the continuum, through the use of variational techniques, skein
relations have been provided for the Kauffman bracket knot polynomial
\cite{GaPu96}. These results contain as a particular case skein
relations for the linking number with intersections. It is worthwhile
pointing out that these results agree with the results we presented in
the previous paragraphs. The skein relation found states that,
\begin{equation}
N(\raisebox{-3pt}{\psfig{figure=li.eps,height=4mm}}) = {1\over 2}(
N(\raisebox{-3pt}{\psfig{figure=lplus.eps,height=4mm}})+
N(\raisebox{-3pt}{\psfig{figure=lminus.eps,height=4mm}}))
\end{equation}
which implies that the value of the linking number of two loops with an
intersection is equal to average of the values of the linking number
when that intersection is replaced by an upper or an under crossing. In
one case the invariant will be $1$ and in the other zero and therefore
we recover the lattice result that the linking number of two loops that
are only linked through an intersection is one half. If the intersection
had kinks in it, the continuum results contain a free parameter. This
parameter can be adjusted to agree with the lattice results.

\section{The Hamiltonian constraint}
\subsection{Definition}

\begin{figure}[t]
\hskip 1cm \psfig{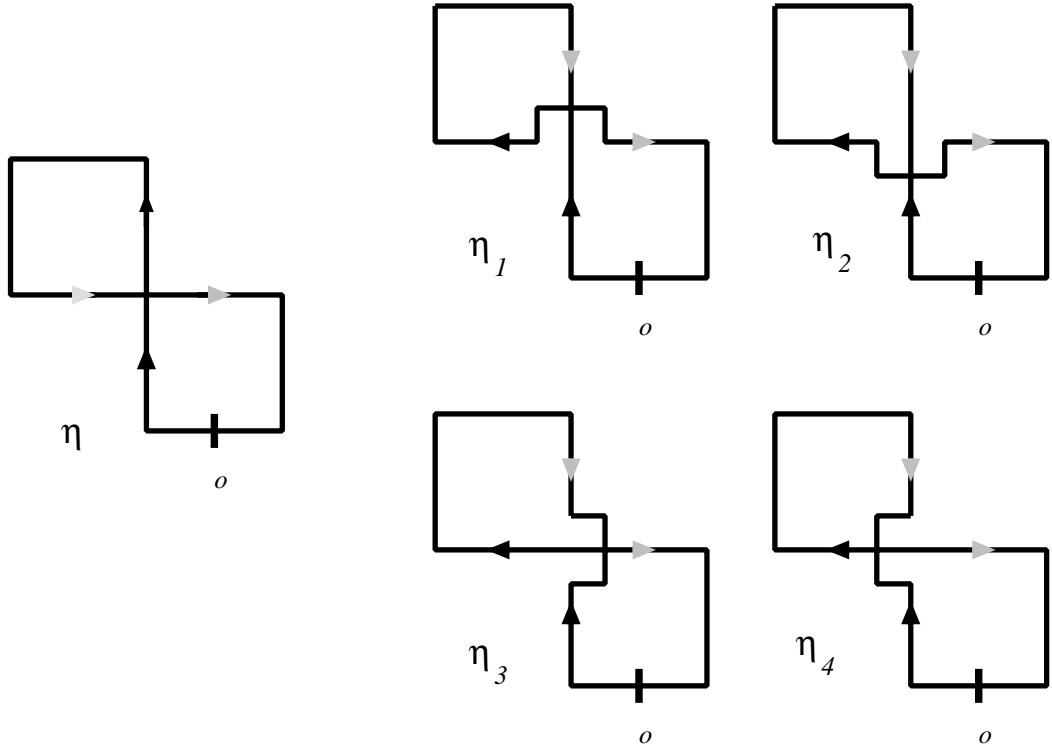}
\caption{The loops that arise in the action of the Hamiltonian on 
a figure eight loop.}
\label{fig8}
\end{figure}

Let us now define the action of the Hamiltonian constraint. We will
propose an operator largely based on the experience of the continuum
\cite{Gaplb91}. We know the operator is only nonvanishing at points
where the loops have intersections. For pedagogical reasons we first
write the definition for a concrete simple loop and then we give the
general definition. Consider a figure eight loop $\eta$ as shown in figure
\ref{fig8}. The action of the Hamiltonian on the state is given by,
\begin{equation}
H(n) \psi(\eta) = {1 \over
4}[\psi(\eta_1)-\psi(\eta_2)+\psi(\eta_3)-\psi(\eta_4)].
\end{equation}
The action of the operator at intersections can be described as
deforming the loop along one of the tangents at the intersection and
rerouting one of the resulting lobes minus the rerouted deformation in
the opposite direction. This operation is carried along for each
possible pair of tangents at the intersection. The ``lobes'' do not
need to be single lobes as depicted here, the situation is the same in
multiple intersections, each pair of tangents determines univocally
two lobes in the loop. 

Let us now describe the general definition of the Hamiltonian acting
on a loop with possible multiple intersections at a point with
possible kinks. We start from a generic loop on the lattice described 
by a chain of links,
\begin{equation}
\gamma= (\ldots,u_-(n),u_+(n),\ldots,v_-(n),v_+(n),\ldots)
\end{equation}
where $u_-(n),u_+(n),\ldots$ are the incoming and outgoing links at the
intersection, which we locate at the site $n$. The intersection can 
be multiple, in that case one has to choose pairs of lines that
go through it and add a contribution similar for each of them.

The general action of 
the Hamiltonian is defined as,
\begin{equation}
{\cal H}(n) \psi(\gamma) = {1 \over 8}I(n,\gamma)
\sum_{l(n),l'(n)\in \gamma}
[\psi(\gamma'_{l,o,l'} \circ {\bar\gamma'_{l,l'}})-
\psi(\gamma''_{l,o,l'} \circ {\bar\gamma''_{l,l'}})] \label{hamilt}
\end{equation}
where the loop $\gamma'$ is defined by,
\begin{equation}
\gamma' \equiv (....u_+(n),u_-(n),u_+(n),{\bar
u}_+(n)...u_+(n),v_-(n),v_+(n),{\bar u}_+(n)...)
\end{equation}
\begin{equation}
\gamma'' \equiv (....\bar{u}_+(n),u_-(n),u_+(n),
u_+(n)...{\bar u}_+(n),v_-(n),v_+(n), u_+(n)...)
\end{equation}
where the function $I(n,\gamma)$ is one if $n$ is at an intersection
of $\gamma$ and the summation goes through all pairs of links $l,l'$
that start or 
end at $n$. The notation $\gamma'_{l,o,l'}$ indicates the portion
of the loop $\gamma'$ that goes from the link $l$ to $l'$ through an
arbitrary origin $o$ fixed along the loop, whereas $\gamma'_{l,l'}$
represents the rest of the loop. An overbar denotes reverse
orientation. Therefore $\gamma'_{l,o,l'} \circ {\bar\gamma'_{l,l'}}$
corresponds, in the case of a figure eight loop to the deformed and
rerouted loop we discussed above as $\eta_1$.

An important property of the Hamiltonian that needs to be pointed out
is that its action on double intersections trivializes in the space of
wavefunctions that are invariant under diffeomorphisms. If the
wavefunction is invariant under diffeomorphisms, it is a fact that 
$\psi(\eta_1)=\psi(\eta_2)$ and $\psi(\eta_3)=\psi(\eta_4)$ where
the $\eta_1,\ldots,\eta_4$ refer to the loops in figure
\ref{fig8}. Therefore we need only concern ourselves, when analyzing
solutions in terms of knot invariants, with triple or higher
intersections. In the lattice, if one is not considering double lines
(as we are doing in this paper for simplicity) that means only up to
triple intersections (possibly with kinks).

\subsection{Continuum limit}

Let us analyze the continuum limit of the above expression. What we
would like to study is its action on a holonomy based on a smooth loop
of a smooth connection and show that it yields the same expression as
the action of the Hamiltonian constraint of quantum gravity on a
holonomy in the connection representation. In order to see this let us
evaluate the Hamiltonian we just introduced on a Wilson loop
$W(\gamma) ={\rm Tr}(U(\gamma))$ where $U(\gamma)$ is the holonomy,
\begin{equation}
{\cal H}(n) W(\gamma) = {1 \over 8}I(n,\gamma)
\sum_{l(n),l'(n)\in \gamma}
{\rm Tr}[U(\gamma'_{l,o,l'} \circ {\bar\gamma'_{l,l'}})-
U(\gamma''_{l,o,l'} \circ {\bar\gamma''_{l,l'}})] 
\end{equation}
and as in the section where we discussed the diffeomorphism
constraint, we represent the deformation of the loop introduced by the
Hamiltonian through the insertion of $F_{ab}$'s in the holonomy
multiplied by the element of area of the plaquette, specifically we need 
to compute the deformations of the rerouted loop shown in figure \ref{fig8},
\begin{eqnarray}
{\cal H}(n) W(\gamma) = {1 \over 4}I(n,\gamma)
\sum_{l'(n)<l(n)\in
\gamma}
&\{&
Tr[(1+a^2F_{ab}(\gamma_o^{l'}))(1+a^2F_{ba}(\gamma_o^{l'}
\circ{\bar\gamma_{l}^{l'}}))
 U(\gamma_{l,o,l'} \circ {\bar\gamma_{l,l'}})] \nonumber \\
&&- Tr[(1+a^2F_{ba}(\gamma_o^{l'}))(1+a^2F_{ab}(\gamma_o^{l'}
\circ{\bar\gamma_{l}^{l'}}))
 U(\gamma_{l,o,l'} \circ {\bar\gamma_{l,l'}})]  \nonumber\\
&&+ Tr[(1+a^2F_{ba}(\gamma_o^{l'}\circ{\bar\gamma_{l}^{l'}}))
(1+a^2F_{ab}(\gamma_o^{l'}))
 U(\gamma_{l,o,l'} \circ {\bar\gamma_{l,l'}})]  \nonumber\\
&&- Tr[(1+a^2F_{ba}(\gamma_o^{l'}))(1+a^2F_{ab}(\gamma_o^{l'}
\circ{\bar\gamma_{l}^{l'}}))
 U(\gamma_{l,o,l'} \circ {\bar\gamma_{l,l'}})]\}
  u^a(l) u^b(l')
\end{eqnarray}
so the action of the Hamiltonian, listing explicitly the lattice
spacing,  is finally given by, 
\begin{equation}
{{\cal H}(n) \over a^6} W(\gamma) = a^2 {I(n,\gamma)\over a^6}
\sum_{l'(n)<l(n)\in \gamma}
{\rm Tr}[F_{ba}(n) U(\gamma_{l,o,l'} \circ {\bar\gamma_{l,l'}})] 
u^a(l) u^b(l').
+ {\rm Tr}[F_{ab}(n) U(\bar\gamma_{l,l'} \circ 
{\gamma_{l,o,l'}})] u^a(l) u^b(l').
\end{equation}

This expression is immediately identified as the lattice version of the
continuum formula (see for instance equation 8.32 of \cite{GaPubook}),
\begin{equation}
{\cal H}(x) W(\gamma) = \oint_\gamma dy^a \oint_\gamma dz^b 
\delta(x-y)\delta(x-z)
{\rm Tr}F_{ab}(y)[U(\gamma_{y,o,z} \circ {\bar\gamma_{y,z}})+
U({\bar\gamma_{y,z}} \circ\gamma_{y,o,z})]
\end{equation}
which has been the starting point for the formulation of the
Hamiltonian constraint in the loop representation \cite{Gaplb91}, and
which corresponds to the action of the Hamiltonian constraint in the
connection representation when acting on a Holonomy,
\begin{eqnarray}
{\cal H}(x) W(\gamma) &=&  {\rm Tr}[F_{ab}(x)
{\delta \over \delta A_a(x)} {\delta \over \delta A_b(x)}] U(\gamma)\\
&=& {\rm Tr}( F_{ab}(x)\hat{E}^a(x) \hat{E}^b(x)) W(\gamma).
\end{eqnarray}

There are some general comments that one can make about this particular
choice of Hamiltonian constraint. Of course, there is significant
freedom in defining the operator. The choice we have made here,
especially in what concerns how to implement the action of the curvature
that appears in the constraint, is in line with what we did for the
diffeomorphism constraint. One could suggest immediately other
possibilities. The first that comes to mind is to implement the
curvature instead of shifting the loop to the right and left by elements
of area, to only shift in one direction. This less symmetrical ordering
yields a quite different action of the Hamiltonian. Whereas our present
choice converts triple straight through intersections into double ones,
the unsymmetrical choice maps triple intersections to double and triple
ones. As we will see in the next section this would imply a difference
in the kinds of solutions one would find. It might also impact the issue
of the constraint algebra. Although we will not discuss the constraint
algebra for the Hamiltonian we are proposing, one can see that problems
might arise in computing the commutator of two Hamiltonians
successively, since they tend to reduce the order of intersections. One
should take these comments cautiously. All these issues
require to examine the action of the Hamiltonian on all possible kinds
of intersections (for instance, multiply traversed, or with kinks), 
not just straight through ones.

\section{Solutions to the Hamiltonian constraint}

We will now analyze a solution to the Hamiltonian constraint in terms
of knot invariants on the lattice, which therefore are solutions of
the diffeomorphism constraint as well and as a consequence are quantum
states of gravity. It has been known for some time from 
formal calculations in  the continuum
\cite{BrGaPu} that the second coefficient of the Conway polynomial is
annihilated by the Hamiltonian constraint of quantum gravity. This was
concluded after lengthy formal calculations both in the loop
representation and in the extended loop representation
\cite{DiGaGr}. In the latter case a regularized calculation was
performed but it required the introduction of counterterms. We will
show here that the lattice version of the second coefficient of the
Conway polynomial is annihilated by the lattice Hamiltonian
constraint. 

We will also show that the third coefficient in terms of $\Lambda$ in
the $q=e^\Lambda$ expansion of the Jones polynomial is annihilated by
the Hamiltonian constraint. Prima facie, this {\em does not} agree
with the results of the extended loop representation. Although the
terms generated by the action of the Hamiltonian are quite similar in
both cases, they are not exactly equal and a cancellation occurs in
the lattice that does not happen in terms of extended loops. This
could be ascribed simply to the fact that we are dealing with two
different regularizations and therefore the results do not necessarily
have to agree or to the fact that in the lattice we are exploring only
a few examples of possible intersections while doing the calculations
and therefore it could simply be that for more
complex intersections it is not annihilated and the results {\em do}
agree with those of the extended loop representation. The extended
loops naturally incorporate all possible types of intersections from
the outset.

It is worthwhile recapitulating a bit on the results of the extended loop
approach to give a context to the solutions we will present in the
lattice, although the derivations are completely independent. It is a
known fact that the exponential of the Chern--Simons form is a solution
of the Hamiltonian constraint with a cosmological constant $\Lambda$ in
the connection representation \cite{Ko,BrGaPu},
\begin{equation}
\Psi^{CS}[A] = \exp\left(-{6\over \Lambda} S_{CS}\right)
\end{equation}
where $S_{CS}= \int d^3 x {\rm Tr}[A \wedge \partial A +{2\over
3} A\wedge A \wedge A]$. It is not difficult to see that this state
is annihilated by the Hamiltonian constraint with cosmological constant.
This state has the property that the action of the magnetic field 
quantum operator on it is proportional to the electric field (in the
gravity case the triad) operator. That makes the two terms in the
Hamiltonian constraint proportional to each other. One adjusts the
proportionality factor so they cancel.

If one transforms this state into the loop representation one gets,
\begin{equation}
\Psi^{CS}(\gamma)= \int DA\, e^{{6\over \Lambda} S_{CS}} W_\gamma[A]
\end{equation}
so this is equivalent to the expectation value of the Wilson loop in a
Chern--Simons theory. This is a well studied quantity first considered by
Witten \cite{Wi89} and it is known
that for the case of $SU(2)$ connections that 
it corresponds to a knot polynomial that is
known as the Kauffman bracket with the polynomial variable $q$ taking the 
value $q=\exp \Lambda$. It is also well known that the Kauffman bracket
is related in the vertical framing to the Jones polynomial through the
relation,
\begin{equation}
{\rm Kauffman}\, {\rm Bracket}_{e^\Lambda}(\gamma)=\exp(\Lambda {\rm
Gauss}(\gamma))\, {\rm Jones}_{e^\Lambda}
\end{equation}
where Gauss$(\gamma)$ is the self-linking number of the loop $\gamma$.

Using first formal loop techniques in the continuum and later extended
loop techniques, it was established that in order for the Kauffman
bracket to be a solution of the Hamiltonian constraint with
cosmological constant, it had to happen that the second coefficient of
the Jones polynomial (not in its original variable, but in the
infinite expansion resulting of setting it equal to $\exp(\Lambda)$)
had to be annihilated by the Hamiltonian constraint of quantum gravity
(with no cosmological constant). The coefficients of the Jones
polynomial in this infinite expansion are known to be Vassiliev
invariants and the second one is known to coincide up to numerical
factors with the second coefficient of the Alexander--Conway \cite{Co}
knot polynomial. The third coefficient in the expansion was shown {\em
not} to be a solution of the Hamiltonian constraint (with or without
cosmological constant) \cite{Gr}.

We will now examine the counterpart of these results in the lattice
approach. Of course, a full examination of any solution would require
evaluating the action of the Hamiltonian on all possible types of
intersections, including multiple lines. We will not do this here. We
will first study triple straight-through intersections and we will end
the section with a discussion of kinks.

\subsection{The second coefficient}

We need to define the second coefficient of the Alexander--Conway \cite{Co}
polynomial in the lattice, including intersecting loops. In order to do
this we draw from our knowledge of the fact that the second coefficient
of the Alexander--Conway polynomial coincides with the second
coefficient of the expansion of the Jones polynomial in terms of
$\Lambda$ when the polynomial variable is $\exp(\Lambda)$. As we
discussed before, this is related to the expectation value of a Wilson
loop in a 
Chern--Simons theory. Putting together the results of \cite{GaPu96} with
the discussion of Abelian Chern--Simons theory on the lattice we presented
here (whic allows us to find the Gauss linking number with
intersections) one gets that the second coefficient $a_2(\gamma)$
satisfies, 
\begin{eqnarray}
a_2(\raisebox{-3pt}{\psfig{figure=lplus.eps,height=4mm}}) -
a_2(\raisebox{-3pt}{\psfig{figure=lminus.eps,height=4mm}}) &=&
a_1(\raisebox{-3pt}{\psfig{figure=l0.eps,height=4mm}})\\
a_2(\raisebox{-3pt}{\psfig{figure=lo.eps,height=4mm}}) &=& 0, 
\end{eqnarray} 
where $a_1$ is an invariant that coincides with the Gauss linking
number if the involved loop has two components 
(up to a factor, actually $a_1(\gamma_1,\gamma_2)=3
lk(\gamma_1,\gamma_2)$ with the definitions of this paper) 
and is zero otherwise. 

The relationship to the 
Chern--Simons state allows us to define an extension of the second
coefficient for intersections, using the techniques of 
reference \cite{GaPu96}.  The result is, 
\begin{eqnarray}
a_2(\raisebox{-3pt}{\psfig{figure=li.eps,height=4mm}}) &=& {1\over 2}
(a_2(\raisebox{-3pt}{\psfig{figure=lplus.eps,height=4mm}}) +
a_2(\raisebox{-3pt}{\psfig{figure=lminus.eps,height=4mm}})) +{1\over 8}
(a_0(\raisebox{-3pt}{\psfig{figure=lplus.eps,height=4mm}}) +
a_0(\raisebox{-3pt}{\psfig{figure=lminus.eps,height=4mm}}))\\
a_2(\raisebox{-3pt}{\psfig{figure=lv.eps,height=4mm}}) &=&
\label{pricecol} a_2(\raisebox{-3pt}{\psfig{figure=l0.eps,height=4mm}})
\end{eqnarray} where $a_0=2^{n_c}/2$ where $n_c$ is the number of
connected components of the knot ($a_0=1$ for a single component knot
.)  

The above relations imply that for the second coefficient we can turn
an intersection into an upper or under-crossing ``at the price'' of a
term proportional to the linking number and another one to the 
number of connected components,
\begin{eqnarray}
a_2(\raisebox{-3pt}{\psfig{figure=li.eps,height=4mm}}) &=& 
a_2(\raisebox{-3pt}{\psfig{figure=lminus.eps,height=4mm}}) +
{1\over 2} a_1(\raisebox{-3pt}{\psfig{figure=l0.eps,height=4mm}}) 
+{1\over 4}a_0(\raisebox{-3pt}{\psfig{figure=lminus.eps,height=4mm}}) 
\label{priceup}\\
&=&a_2(\raisebox{-3pt}{\psfig{figure=lplus.eps,height=4mm}}) -
{1\over 2} a_1(\raisebox{-3pt}{\psfig{figure=l0.eps,height=4mm}})
+{1\over 4}a_0(\raisebox{-3pt}{\psfig{figure=lplus.eps,height=4mm}}). 
\label{pricedown}
\end{eqnarray}

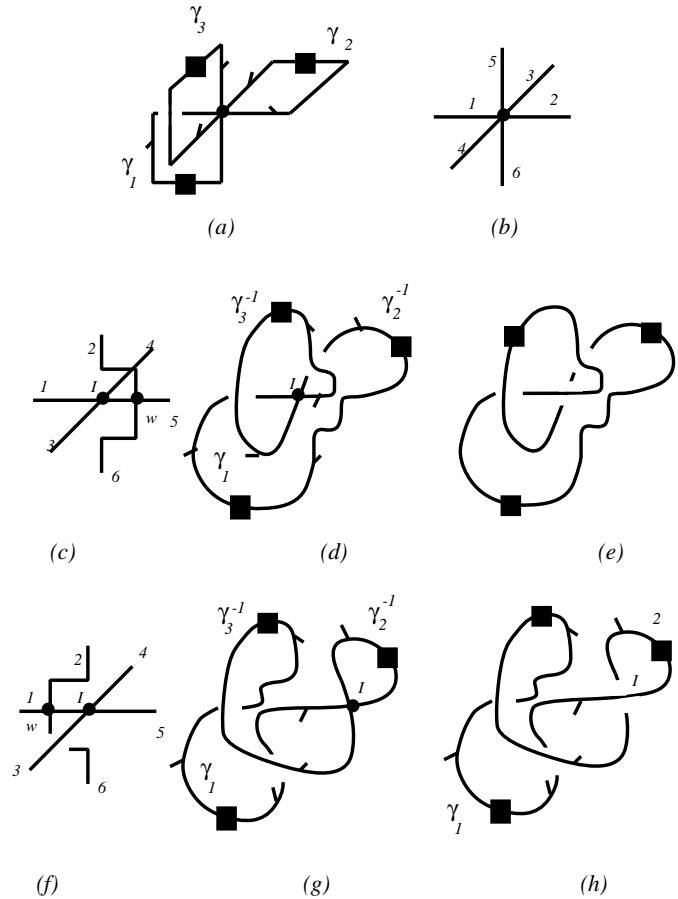
\begin{figure}[b]
\hspace{3cm}{\psfig{figure=fig4.eps,height=12cm}}
\vspace{0.1cm}
\caption{The action of the Hamiltonian on the generic loop considered
in the $a_2$ calculation. Given a generic loop with a triple
intersection (a), the Hamiltonian splits the triple intersection (b)
into two double ones, back (f) and forth (c). For the case of the
second coefficient of the Conway polynomial the resulting loop can be
rearranged into a loop without intersections using the skein
relations. First one uses the skein relation for an intersection with
a kink and obtains loop (d).  Then one uses the skein relation for
regular intersections to convert the loop to a loop without
intersections. Similar operations are depicted in (f-h) for the other
deformation produced by the action of the Hamiltonian. From (f) to (g)
one again uses the skein relation for the intersection with a kink and
from (g) to (h) one uses the one for regular intersections.  Since the
two resulting (h) and (d) are deformable to each other the
contributions from the $a_2$ cancel. One is left with the
contributions of linking numbers and connected components introduced
when one uses the skein relation for regular intersections. When one
considers these contributions for all possible pairs of tangents the
resulting expression vanishes taking into account the Abelian nature
of $a_1$ and $a_0$ . The black squares indicate that the different
lobes of the loop could have arbitrary knottings and
interlinkings. The result only depends on the local connectivity at
the intersection.}
\label{a2fig}
\end{figure}

We now consider the action of the Hamiltonian on $a_2$.  We start with
a straight-through intersection for simplicity, we will later discuss
the case of an intersection with a kink.  The action of the
Hamiltonian constraint requires considering the six possible pairs of
tangents that are involved in a triple intersection and in each case
produces two terms, deforming back and forth along one tangent the
loop in the direction given by the other tangent.  In the case of the
straight through intersection, due to symmetry reasons, it is only
necessary to consider three pairs of tangents, the other three yield
the same contributions.  The action of the Hamiltonian is easily
understood pictorially, so we refer the reader to figure \ref{a2fig}.
As we see, the orientation and routing through the intersection of the
original loop is labelled by the numbers 1-6 at the intersection.  The
only information needed from the loop is that it has 
a triple intersection and the connectivity suggested
by the numbers, otherwise it can have any possible kind of knottings
or interlinkings between petals 
away from the intersection, we denote this in the
figure with black boxes.

As can be seen in the figure, the action of the Hamiltonian at the
first pair of tangents we choose (2 and 5) produces as a result the
difference of value of the second coefficients evaluated for two
different loops.  We will now study the value of this difference using
the skein relations for the second coefficient.  Looking at figure
\ref{a2fig} we notice that in the loop obtained deforming to the
right, one of the intersections we are left with (the one at the
right) is a ``collision type'' intersection (marked $w$ in (c)) and
using the skein relation (\ref{pricecol}) we can remove the
intersection simply separating the lines.  Similar comments apply to
the intersection at the left in the loop obtained deforming the
original triple intersection to the left (f).  We are now left with
the second coefficient evaluated on two different loops, each of them
with a single, straight through intersection (figures \ref{a2fig}d,g).
We can remove that intersection either using (\ref{priceup}) or
(\ref{pricedown}).  The remarkable thing is that if we apply
(\ref{priceup}) in the diagram \ref{a2fig}d (``lifting the line") and
(\ref{pricedown}) in \ref{a2fig}g (``lowering the line") we produce
two loops with exactly the same topology.  To be more precise,
formulas (\ref{priceup},\ref{pricedown}) have two types of
contributions, $a_2$ and $a_0$ evaluated on the loop with the line
raised or lowered, plus $a_1$ contributions.  The point is that by
arranging things in the way we did, we end up with the difference of
the second coefficients evaluated on loops with exactly the same
topology (the difference comes because the ``forward'' and
``backward'' actions in the Hamiltonian come with opposite signs).
Therefore this contribution cancels (so does the contribution of the
$a_0$'s).  Let us now concentrate on the terms involving $a_1$'s.
These terms imply replacing the intersection by a reconnection
of the original loop into two disjoint loops. For figure
\ref{a2fig}d this yields a link consisting of $\gamma_3^{-1}$ and
$\gamma_1\circ \gamma_2^{-1}$. Given that $a_1$ for links with two
components is (up to a factor of 3 that is irrelevant for what
follows) the linking number of the components, the result for this
contribution is,
\begin{equation}
lk(\gamma_3^{-1},\gamma_1\circ\gamma_2^{-1})=
-lk(\gamma_3,\gamma_1)+lk(\gamma_3,\gamma_2)
\end{equation}
due to the Abelian relations (\ref{abel1},\ref{abel2}). For the loop 
\ref{a2fig}g the linking number contribution is
\begin{equation}
lk(\gamma_2^{-1},\gamma_3^{-1}\circ\gamma_1) = 
lk(\gamma_2,\gamma_3)-lk(\gamma_2,\gamma_1)
\end{equation}
so the total contribution to the Hamiltonian due to the deformations
along the (2-5) tangents is equal to
\begin{equation}
-lk(\gamma_3,\gamma_1)+lk(\gamma_2,\gamma_1).\label{1cont}
\end{equation} 

\begin{figure}[b] 
\hspace{3cm}{\psfig{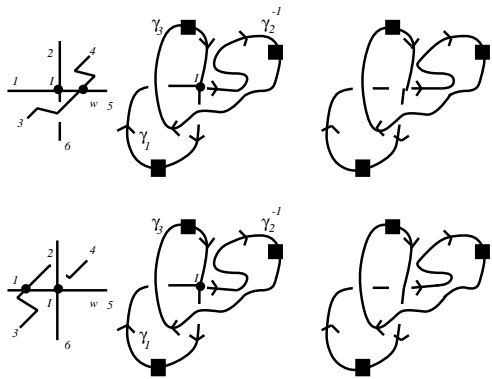}}
\caption{The contribution to the Hamiltonian of the
deformation along the (2-3) pair of tangents.} \label{a2fig2}
\end{figure}

\begin{figure}[b]
\hspace{3cm}{\psfig{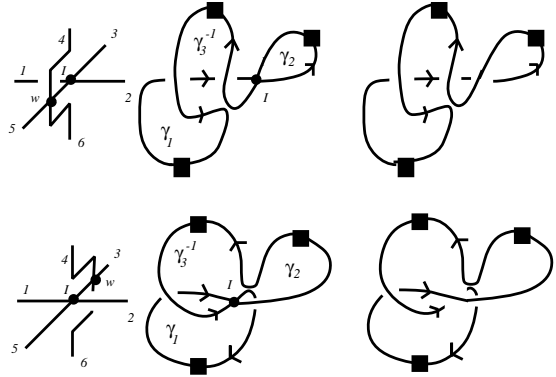}}
\caption{The contribution to the Hamiltonian of the deformation along
the (5-4) pair of tangents.}
\label{a2fig3}
\end{figure}

We now study the deformations along the (2-3) and (5-4) pairs of
tangents (the labels refer to the original intersection, shown in
figure \ref{a2fig}a), as shown in figures \ref{a2fig2} and
\ref{a2fig3}. Using exactly the same ideas, ie, using (\ref{pricecol})
to remove collision intersections and (\ref{priceup}) or
(\ref{pricedown}) in straight through intersections in such a way as
to cancel all $a_2$ and $a_0$ contributions we get for the deformation
in figure \ref{a2fig2},
\begin{equation}
lk(\gamma_3\circ\gamma_2^{-1},\gamma_1)-
lk(\gamma_3,\gamma_2^{-1}\circ\gamma_1)=
-lk(\gamma_1,\gamma_2)+lk(\gamma_3,\gamma_2)\label{2cont}
\end{equation}
and similarly for the contribution along (5-4),
\begin{equation}
lk(\gamma_2,\gamma_3^{-1}\circ\gamma_1)-
lk(\gamma_3^{-1}\circ\gamma_2,\gamma_1)=
-lk(\gamma_2,\gamma_3)+lk(\gamma_1,\gamma_3) \label{3cont}
\end{equation}
and we therefore see, by adding (\ref{1cont},\ref{2cont},\ref{3cont})
that the total contribution to the Hamiltonian vanishes.

Rigorously one should carry out this kind of computations 
for {\em all} possible pairs of
tangents. It turns out that due to symmetry reasons, for a straight
through intersection the other three pairs produce exactly the same
contributions as the ones we listed here. We have presented enough
elements for the reader to check this fact if needed.

So we see that the root of the annihilation of the Hamiltonian
constraint on the $a_2$ can be traced back to the fact that the skein
relations for that coefficient relate it to the $a_1$, which is a very
simple invariant given its ``Abelian" nature and that allows for several
cancellations to occur. We will see  that rather unexpectedly 
this seems to be also
the case for higher coefficients.

\subsection{The third coefficient}

The third coefficient in the expansion in terms of $q=e^\Lambda$ of
the Jones polynomial is not annihilated by the Hamiltonian constraint
in the extended loop representation \cite{Gr}.  We will here apply our
lattice Hamiltonian to the third coefficient.  We will see that it is
annihilated by the lattice Hamiltonian, at least for straight-through
triple intersections.  The calculation goes exactly along the same
lines as in the previous section, that is, the Hamiltonian deforms and
reroutes in the same way.  What changes are the skein relations
involved in managing the resulting action.  Prima facie it seems like
the calculation is not going to give zero. The skein relations for the
third coefficient do not allow us to relate everything to the linking
number and its convenient Abelian properties that helped produce the
cancellation in the $a_2$ case. In fact, this is exactly what prevents
the extended loop calculation from giving zero \cite{Gr}. We will see,
however, that unexpected cancellations take place and the result is
zero.

The skein relations for the third coefficient are, again, obtained by
drawing on the relationship to the Chern--Simons state and using the
techniques of \cite{GaPu96}. The results are,
\begin{eqnarray}
a_3(\raisebox{-3pt}{\psfig{figure=lplus.eps,height=4mm}})-
a_3(\raisebox{-3pt}{\psfig{figure=lminus.eps,height=4mm}}) &=&
a_2(\raisebox{-3pt}{\psfig{figure=l0.eps,height=4mm}})-
a_2(\raisebox{-3pt}{\psfig{figure=lplus.eps,height=4mm}})-
a_2(\raisebox{-3pt}{\psfig{figure=lminus.eps,height=4mm}})-
{1\over 4}\\
a_3(\raisebox{-3pt}{\psfig{figure=li.eps,height=4mm}}) &=&
{3\over 8} a_1(\raisebox{-3pt}{\psfig{figure=l0.eps,height=4mm}})
+a_3(\raisebox{-3pt}{\psfig{figure=lplus.eps,height=4mm}})+
{a_2(\raisebox{-3pt}{\psfig{figure=lplus.eps,height=4mm}})+
a_2(\raisebox{-3pt}{\psfig{figure=lminus.eps,height=4mm}})-
a_2(\raisebox{-3pt}{\psfig{figure=l0.eps,height=4mm}})\over 2}
+{1 \over 8}\\
&=&
{3\over 8} a_1(\raisebox{-3pt}{\psfig{figure=l0.eps,height=4mm}})
+a_3(\raisebox{-3pt}{\psfig{figure=lminus.eps,height=4mm}})-
{a_2(\raisebox{-3pt}{\psfig{figure=lplus.eps,height=4mm}})+
a_2(\raisebox{-3pt}{\psfig{figure=lminus.eps,height=4mm}})-
a_2(\raisebox{-3pt}{\psfig{figure=l0.eps,height=4mm}})\over 2}
-{1 \over 8}\\
a_3(\raisebox{-3pt}{\psfig{figure=lv.eps,height=4mm}}) &=&
a_3(\raisebox{-3pt}{\psfig{figure=l0.eps,height=4mm}})
\end{eqnarray}
where the last three equations show that again one can eliminate
intersections ``at a price''. In this case the price is getting
contributions proportional to the linking number and the second
coefficient. Notice however, that what appears is the second
coefficient evaluated for
$(\raisebox{-3pt}{\psfig{figure=l0.eps,height=4mm}})$. That means that
if one started from a loop with a single component, the loop on which
the second coefficient is evaluated might have two components. Up to
now we had only worried about $a_2$ evaluated for single loops. In
order to know its value for double loops, one can recourse again to
the fact that $a_2$ is related to the Kauffman bracket and the latter
has to satisfy the Mandelstam identities for multiloops. Using the
relation of $a_2$ with the Kauffman bracket, one gets,
\begin{equation}
a_2(\gamma_1,\gamma_2) = a_2(\gamma_1\circ\gamma_2) +
a_2(\gamma_1\circ\gamma_2^{-1}) +{9 \over 2} lk(\gamma_1,\gamma_2)^2,
\label{mandela2}
\end{equation}
where $a_2(\gamma_1,\gamma_2)$ means evaluating the second coefficient
on the link formed by loops $\gamma_1$ and $\gamma_2$.

Armed with these relations, we again go through the same calculation as
before, ie, figures \ref{a2fig}-\ref{a2fig3}. Again eliminating the
$a_3$ contributions by deforming the straight through intersections up 
and down as needed in the back and forth action of the constraint, one
is left with contributions involving $a_2$, $a_1$ and $a_0$. It is not
too lengthy to prove that these contributions cancel and one is only 
left with contributions of $a_2$'s evaluated on two loops. The result
is,
\begin{eqnarray}
{\cal H} a_3(\gamma_1\circ\gamma_2\circ\gamma_3) &=&
{1\over 8} \left [ 
a_2(\gamma_2^{-1},\gamma_3^{-1}\circ\gamma_1)-
a_2(\gamma_3^{-1},\gamma_2^{-1}\circ\gamma_1)+
a_2(\gamma_3,\gamma_2^{-1}\circ\gamma_1)\right.\\
&&\left. -a_2(\gamma_3,\gamma_2^{-1}\circ\gamma_1)+
a_2(\gamma_1,\gamma_2\circ\gamma_3^{-1})-
a_2(\gamma_2,\gamma_3^{-1}\circ\gamma_1)\right]\nonumber
\end{eqnarray}
where we have denoted by $\gamma_1$, $\gamma_2$ and $\gamma_3$ the
three ``petals'' determined by the triple self-intersection of the
loop on which we are acting. We also denote by $\circ$ the composition
of loops as before.  

If one now uses Eq. (\ref{mandela2}) and the Abelian properties of 
the linking number one sees that all contributions involving linking
numbers cancel. Furthermore, using that the remaining Mandelstam
identities for $a_2$, namely 
\begin{eqnarray}
a_2(\gamma)&=&a_2(\gamma^{-1})\\
a_2(\gamma_1\circ\gamma_2\circ\gamma_3)&=& 
a_2(\gamma_2\circ\gamma_3\circ\gamma_1)= 
a_2(\gamma_3\circ\gamma_2\circ\gamma_1)
\end{eqnarray}
one sees that all contributions finally cancel.

It should be remarked that the $a_3$ could not  be a state of quantum
gravity even if it is annihilated by the Hamiltonian constraint because
it does not satisfy the Mandelstam identities that loop states have to
satisfy. This can be easily seen again by considering the fact that 
the Kauffman bracket does satisfy it and at third order in the 
expansion the coefficient of the Kauffman bracket is given by a 
combination involving $a_3$ plust $a_2$ times the linking number and
the linking number cube. It is easy to see that $a_2$ times the
linking number does not satisfy the Mandelstam identities, so $a_3$ 
does not.

Therefore we see that for this particular kind of loops the result
vanishes in a very nontrivial way. It is yet to be determined what
will happen for loops with kinks or multiple lines at intersections.

\subsection{Intersections with kinks}

\begin{figure}
\hspace{3cm}{\psfig{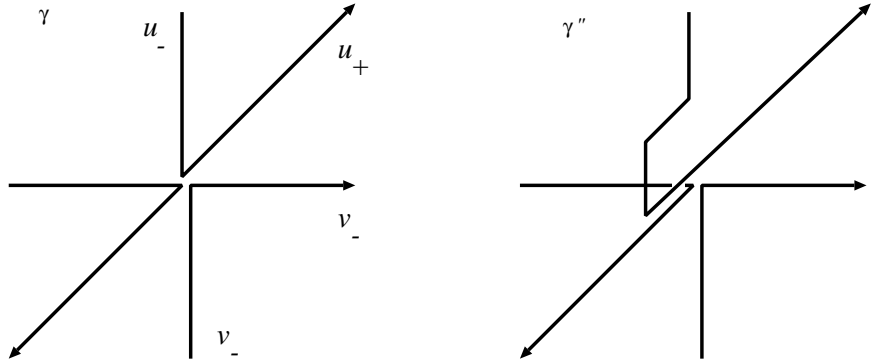}}
\label{intkink}
\caption{The action of the Hamiltonian on a triple intersection with
kinks is not simply given by double intersecting loops.}
\end{figure}

The action of the Hamiltonian we have is well defined at intersections
with kinks, in the sense that it is immediate to see that it has the
correct continuum limit.  Therefore one is in a position to attempt to
apply the Hamiltonian to the coefficients we studied in the previous
sections for the case with kinks.  In the lattice there is only a
finite number of possible cases ---at least if one restricts oneself
to simply traversed lines--- so there is the chance of carrying out an
exhaustive analysis.  Rigorously speaking it is not true that a
wavefunction is a state of quantum gravity until one has completed
such analysis.  We will not, however, carry out this analysis here.
The reason for this is that the Hamiltonian acting on a triple
intersection with kinks (it is easy to see that for double
intersections it again identically vanishes) has a more complex action
than on straight through intersections.  In straight through triple
intersections the Hamiltonian produced a loop with double
intersections only.  This is not the case if there are kinks, as
figure \ref{intkink} shows.  This is actually quite reasonable.  If
the action of the Hamiltonian always simplified the kind of
intersection one runs the danger of producing an operator that would
identically vanish after successive applications.  This is obviously
not reasonable for the Hamiltonian constraint.  It would also pose
immediate problems for the issue of the constraint algebra. However,
this reasonable fact complicates significantly the application of the
Hamiltonian to the coefficients we discussed before in the case of
kinks, since we now need skein relations for the coefficients, not
only for triple intersections, but for triple intersections with kinks
and double lines. The techniques of \cite{GaPu96} actually allow to
figure out such skein relations, but a detailed analysis has not yet
been performed. Such an analysis is more complex than that of double
intersections, since it was shown that there are regularization
ambiguities in the definition of skein relations for knot invariants
with kinks. The calculations involved, however, are well defined and
the analysis can be performed, it just exceeds the scope of this
paper.

The lack of a confirmation that the Hamiltonian annihilates the
coefficients in the case of kinks prevents us from really claiming
that one has a rigorous proof that the coefficients are solutions. In
fact, it could well be that the third coefficient is not annihilated
in the case of intersections with kinks and therefore the results
presented are in fact compatible with the extended loop
results. Further investigations should confirm if this is the case.

Even if we were able to perform the calculations for all the possible
cases with kinks, we should remind that we are only dealing here with
simply-traversed loops. For multiply-traversed loops again the
computations are well defined, but have not been performed. In fact,
the number of possible cases to be analyzed is infinite, since one has
to consider multiple lines. One might worry therefore that it will
never be possible to show that a state is a solution. It might be
useful to rethink all of this approach in terms of spin networks,
where a more unified treatment of all possible cases can be had from
the outset. It might be possible that generic behaviors are found and
one is not really required to compute all possible cases.

\subsection{Regular isotopic invariants}

In the extended loop representation, apart from the second and third
coefficient, there exist other states of quantum gravity if one adds a
cosmological constant to the theory.  The addition of a cosmological
constant implies that the Hamiltonian constraint gains a terms of the
form $\Lambda {\rm det} g$ proportional to the determinant of the
three dimensional metric.  It is not difficult to implement such a
term in the lattice.  Based on its action in the continuum
\cite{GaPuBa}, which basically acts nontrivially only at triple
intersections and introduces reroutings and kinks (no
deformations). One would therefore be tempted to try to apply the
Hamiltonian with cosmological constant in the lattice on these states
and check if it is satisfied. Unfortunately, all states with
cosmological constant that appear in the extended loop representation
(essentially the exponential of the self-linking number and the
Kauffman bracket polynomial itself) are framing-dependent
invariants. This is not a problem in the extended representation,
which in a sense is a framing procedure, but is an inescapable
obstacle in the lattice context. As we argued in the section on knot
theory on the lattice, although lattice Chern--Simons theory provides
an unambiguous definition for the framing dependent invariants, the
resulting objects are not annihilated by the diffeomorphism
constraint.  Therefore the most reasonable attitude is to not consider
these states in this context, at least in the particular
implementation we are proposing.

\section{Conclusions}

This paper can only be taken as the first exploratory steps in the
construction of a possible theory of quantum gravity using lattice
regularizations. We have discussed several issues we would like to
summarize here.

a) It seems possible to implement a lattice notion of
``diffeomorphisms'' in the sense of having a set of symmetries that
become diffeomorphisms in the continuum.  These operations are
represented quantum mechanically by an operator that has the correct
continuum limit and the correct algebra in the continuum limit.
Moreover, the solutions to the constraint become in the continuum knot
invariants and are the basis for establishing a notion of knot theory on
the lattice.

b) We proposed a lattice regularization of the Hamiltonian constraint
that has a simple and geometrical action on the lattice. In fact, its
action can be viewed as a set of skein relations in knot space, though
we leave the discussion of this viewpoint to a separate paper
\cite{GaPu96b}. 

c) We tie up together the results of knot theory in the lattice and the
Hamiltonian constraint by showing that certain knot invariants are
actually annihilated by the constraint (at least in certain examples of
loops). These results have a direct counterpart in the continuum. They
also exhibit the strengths and weaknesses of the whole approach: though
we have always well defined computations, to prove results rigorously
(like to claim that a state is annihilated by the constraint), implies
in principle performing an infinite number of computations. 

d) By starting directly in the loop representation and building a theory
that has certain properties that are desirable, one avoids the many 
ambiguities one is faced with if one tries to first implement 
classical general relativity on a lattice, quantize that theory and
only then introduce loops.

Concerning possible future developments of this whole approach, there
are several evident directions that need further study:

a) The issue of the constraint algebra, in particular implying the
Hamiltonian constraint is yet to be completed.

b) The extension of the results to intersections with kinks is needed,
especially to try to settle the issue of the third coefficient and to
consolidate the status of the solutions in general.

c) One could consider extending the results of lattice Chern--Simons 
theory to the non-Abelian case. This would open up the possibility of
applying numerical techniques to the evaluation of the path integrals
involved in the definition of knot invariants. One could use this method
to confirm the results from other approaches. This would require,
however, the development of techniques to handle the
non-positive-definite Chern--Simons action numerically, although some
results in this direction already exist \cite{Gock}.

As far as relations with other approaches that are currently being
pursued in quantum gravity, it is evident that the whole idea of a
lattice approach can benefit many efforts. The details, of course
would change from one particular situation to another. In general it
would be quite important to extend the results presented here to the
spin network context, since it has many attractive features for
dealing with wavefunctions in the loop representation. One problem is
that in spin networks one is dealing from the beginning with arbitrary
number of kinks, multiple lines and intersections, so the whole
approach may have to shift from the one we have taken in this paper,
where we treated all these issues one at a time in increasing
complexity. On the other hand, the payoff could be important: it may
happen that the spin network perspective actually allows to deal with
all these issues simultaneously in an efficient way. For instance
Kauffman and Lins \cite{KaLi} have made important progress in dealing
with certain aspects of intersecting knot theory in terms of the spin
network language. Moreover, recently Thiemann \cite{Th} has introduced
a very promising version of the Hamiltonian constraint of Lorentzian
general relativity (in this paper, as is usual in new variables
formulation, we were really dealing with either the Euclidean theory
or complex general relativity) at a classical level, that can be
promoted to a well defined quantum operator. This is most naturally
accomplished in the context of spin networks. It would be very
interesting if one can find a simple action for the quantum lattice
version Thiemann's Hamiltonian that annihilated the knot polynomials
we discussed in this paper. The techniques we discussed here could be
used as a guideline to perform that calculation.

An interesting possibility would be to try to derive the current
lattice approach from a four dimensional lattice theory using 
transfer matrix methods \cite{Sch64}. This could allow the use of statistical
methods to provide solutions to the Hamiltonian constraint. The
connection with the four dimensional theory should elucidate the
role of the Lagrange multipliers in the lattice theory. As was 
noted in \cite{FrJa} if one starts from a four dimensional lattice 
approach one does not necessarily recover a three dimensional theory
with free Lagrange multipliers, as the one we considered here.

Summarizing, we have shown that a lattice approach to canonical quantum
gravity in the loop representation is feasible, that it allows to pose
several questions of the continuum theory in a well defined setting and
that several results of the continuum are recovered in a rather
unexpected way bringing together notions of knot theory into the lattice
framework. We expect that in the near future several of the developments
outlined above can be pushed forward towards a more complete picture of
the theory.

\acknowledgements

We wish to thank Jorge Griego for help with the $a_3$ calculations.
RG and JP wish to thank Peter Aichelburg and Abhay Asthekar for
hospitality and financial support at the International Erwin
Schr\"odinger Institute (ESI) in Vienna, where part of this work was
completed.  HF wishes to thank Abhay Ashtekar and the CGPG for
hospitality. This work was also supported in part by grants
NSF-INT-9406269, NSF-PHY-9423950, by funds of the Pennsylvania State
University and its Office for Minority Faculty Development, and the
Eberly Family Research Fund at Penn State.  We also acknowledge
support of CONICYT and PEDECIBA (Uruguay).  JP also acknowledges
support from the Alfred P.  Sloan Foundation through an Alfred P.
Sloan fellowship.

\end{document}